\documentclass[11pt,twoside]{book} 
\usepackage{asp2008n}
\usepackage{times}
\usepackage{lscape}
\usepackage{epsf}

\usepackage{graphicx}
\usepackage{amssymb}
\usepackage{longtable}
\usepackage[figuresright]{rotating}
\usepackage{multirow}

\newcommand{\simless}{\mathbin{\lower 3pt\hbox {$\rlap{\raise 5pt\hbox{$\char'074$}}\mathchar"7218$}}}

\mainmatter

\newlength{\deftabcolsep}
\begin{document}


\title{Low Mass Star Formation in the Taurus-Auriga Clouds}   
\author{Scott J. Kenyon}   
\affil{Smithsonian Astrophysical Observatory, 60 Garden Street \\
Cambridge, MA 02138, USA}    
\author{Mercedes G\'omez}   
\affil{Observatorio Astron\'omico, Universidad Nacional de C\'ordoba \\
Laprida 854, 5000 C\'ordoba, Argentina}
\author{Barbara A. Whitney}   
\affil{Space Science Institute, 4750 Walnut Street, Suite 205 \\
Boulder, CO 80301, USA}

\begin{abstract} 
We review the history and structure of star formation in the
Taurus-Auriga dark clouds. Our discussion includes a summary
of the macroscopic cloud properties, the population of single
and binary pre-main sequence stars, the properties of jets
and outflows, and detailed summaries of selected individual
objects. We include comprehensive tables of dark clouds,
young stars, and jets in the clouds.
\end{abstract}



\section{Overview}

\begin{figure}[!ht]
\centering
\plotone{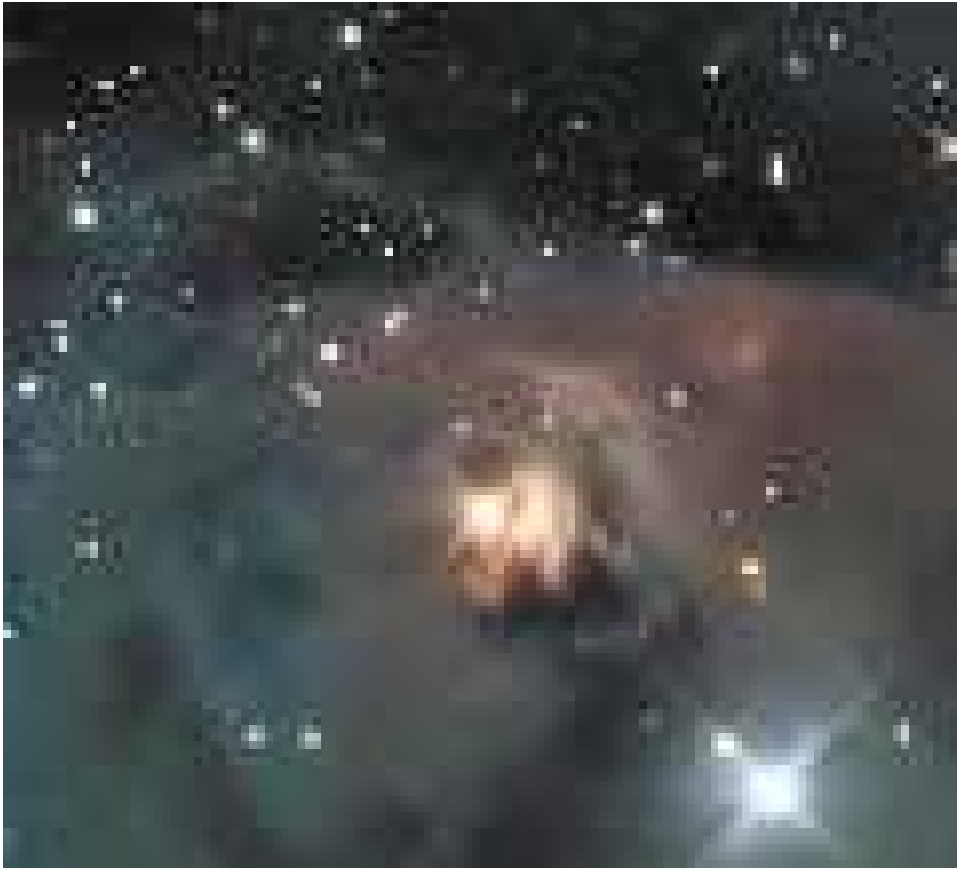}
\caption{Optical image of T Tau and surroundings (courtesy D. Goldman).
North is up and east is to the left.  T Tau is the bright yellow star
near the center. Barnard's nebula is visible as faint nebulosity
immediately surrounding T Tau. Hind's nebula is the bright, arc-shaped
cloud that covers some of the western pair of diffraction spikes from
the T Tau image. Fainter nebulosity, mostly ionized gas powered by
a weak ultraviolet radiation field, covers the rest of the image.
\citet{bur94} and \citet{Bar95} discuss the relationship between
Burnham's nebula and the more distant Hind's and Struve's nebulae.
\label{fig:ttau-opt}
}
\end{figure}

In October 1852, J. R. Hind `noticed a very small nebulous looking
object' roughly 18\arcsec~west of a tenth magnitude star in Taurus.
Over the next 15 years, the nebula slowly faded in brightness and
in 1868 vanished completely from the view of the largest telescopes.
O. Struve then found a new, smaller and fainter, nebulosity roughly
4\arcmin~west of Hind's nebula. While trying to recover these nebulae,
\citet{bur90,bur94} discovered a small elliptical nebula surrounding
T Tau (Figure \ref{fig:ttau-opt}).  Spectra of Hind's nebula revealed
emission from either H$\beta$ or [O~III] $\lambda$5007, demonstrating
that the nebula was gaseous as in novae and planetary nebulae.  At
about the same time, \citet{kno91} reported 4 magnitude variability
in the `ruddy' star associated with these nebulae, T Tauri.

Despite the long history of interpreting variable nebulae
\citep[e.g.,][]{her02,her11}, these discoveries generated
little widespread interest. Among others, \citet{cer06},
\citet{lea07}, and \citet{loc18} identified the unusual
long-period variables RW Aur, UY Aur, RY Tau, and UX Tau on
photographic plates. During the compilation of the HD catalog,
\citet{fle12} noted T Tauri as a long period variable with an
Ma? spectral type. \citet{ada15} and \citet{san20} later
recorded higher quality spectra showing bright emission lines
from H~I and iron. A 5~1/2 hr spectrum taken with the
100\arcsec~reflector revealed many iron absorption lines,
characteristic of late-type stars.

In the 1940's, A. Joy compiled the first lists of `T Tauri stars,'
irregular variable stars associated with dark or bright nebulosity, with
F5-G5 spectral types and low luminosity \citep{joy45,joy49,herbig62}.
Intense searches for other T Tauri stars revealed many stars
associated with dark clouds and bright nebulae, including a class
with A-type spectra \citep[e.g.][]{Her50,her60}. Most of these stars
were in loose groups, the T associations, or in dense clusters, the
O associations \citep[e.g.,][]{herb50,her57,kho58,dol59}.
Because O stars have short lifetimes, both types of associations
have to be composed of young stars, with ages of 10 Myr or less
\citep{amb57}. This realization -- now 50 years old -- initiated
the study of star formation in dark clouds.

\begin{figure}[!ht]
\epsscale{1.0}
\plotone{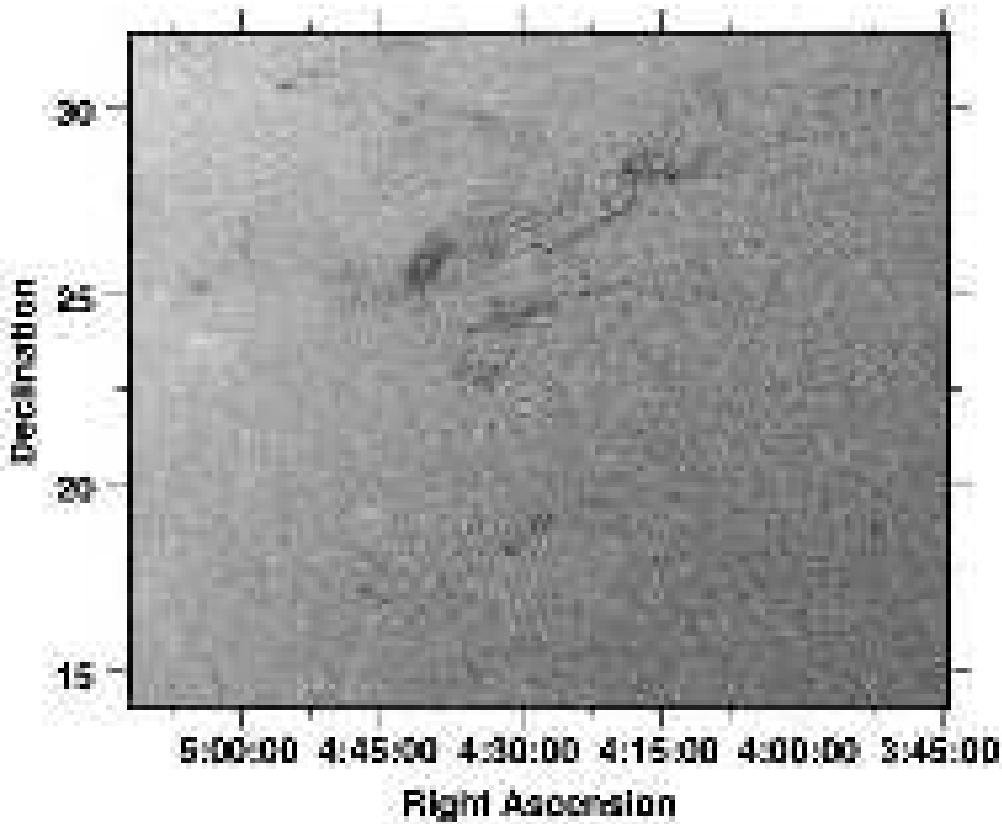}
\caption{Star count map of the Taurus-Auriga dark clouds.  This
map was prepared for this paper from stars with J $\le$ 16.5 in
the 2MASS point source catalog downloaded from the IRSA archive.
The intensity scale is proportional to the number of stars per
square arcmin.  The dark clouds are clearly visible as low density
regions.  Two small bright regions are the open clusters NGC 1647
(RA = 4$^{\rm h}$46$^{\rm m}$, Dec = 19\deg) and NGC 1750/1758
(RA = 5$^{\rm h}$04$^{\rm m}$, Dec = 24\deg).}
\label{fig:counts}
\end{figure}

\section{Cloud Structure}

Shortly after the discovery of variability in T Tauri, Barnard
began to photograph dark nebulae in the plane of the Milky Way
\citep{Bar19, Bar27}. At the time Barnard began this program,
it was still unclear whether dark nebulae were empty regions
of the galaxy or patches of material that obscured background
stars. These photographs showed convincingly that obscuring
material is responsible for many dark clouds. In the B7 cloud,
a partly luminous nebula `seems to fit into a hole in the sky'
\citep{Bar19}. Barnard's photographs showed that many dark clouds
were feebly nebulous, confirming impressions he gained from
visual observations.

Barnard's first catalog of dark nebulae contains 182 objects.  The
Taurus-Auriga clouds comprise roughly a dozen of these dark regions
and lie at a distance of 140--145 pc \citep[Table 1;][]{elias78,
str80,meist81,ken94,wic98,str03,loi05,loi07b,tor07}.
In B7, Barnard noted bright condensations and several small, round
black spots with diameters of 5\arcmin--8\arcmin.  An irregular dark
lane with a width of roughly 10\arcmin~connects B7 to B22. He noted
another dark lane arcing west of B18.

\begin{figure}[tb]
\centering
\plotone{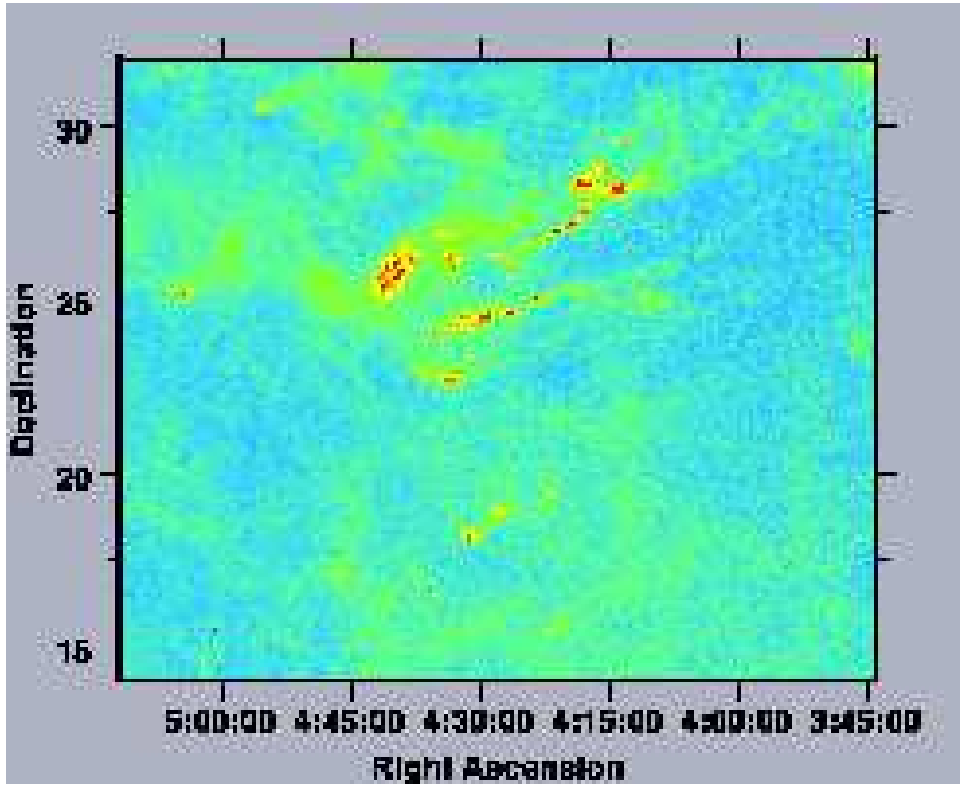}
\caption{Map of the near-infrared color of stars in the Taurus-Auriga
dark clouds. As in Figure \ref{fig:counts}, the map uses stars with
J $\le$ 16.5
from the 2MASS point source catalog.  The intensity scale maps stars
with J--H $<$ 0.5 into blue, stars with J--H = 0.5-1.25 into green,
and stars with J-H $>$ 1.25 into red. The reddest stars lie in or behind
the dark clouds.  Other stars have neutral colors.
}
\label{fig:colors}
\end{figure}

Deeper photographs and digital sky surveys provide the most dramatic
images of the clouds
\citep[Figure \ref{fig:counts}; Table 2;][]{lyn62,Bok77,dob05}.
Barnard's dark spots
and sinuous dark lanes are clearly visible against the bright
background of stars just south of the plane of the Milky Way.
Ultraviolet images show a similar structure \citep{hur94}.
Variations in the ultraviolet (UV) flux across the clouds and counts
of stars in cells yield a good measure of the line-of-sight
density in the cloud; counts in two or more colors allow estimates
of the extinction \citep{str80,meist81,cer84,cer85,cer87,cam99,dob05}.
Improved photometry from digital surveys like 2MASS and the
digitized POSS \citep{dob05} provide more quantitative estimates
of the cloud density.

\begin{table}
\vskip -2ex
\caption{Barnard's Dark Nebulae in Taurus-Auriga}
\smallskip
\begin{center}
{\small
\begin{tabular}[t]{ l l l c | l l l c}
\tableline
\noalign{\smallskip}
ID & $\alpha$(2000) & $\delta$(2000) & D ($^{\prime}$) &
ID & $\alpha$(2000) & $\delta$(2000) & D ($^{\prime}$)\\
\noalign{\smallskip}
\tableline
\noalign{\smallskip}
B7   & 4:17:25 & +28:33	&     & B209 & 4:12:23 & +28:19 \\
B10  & 4:18:41 & +28:16	&  8 & B210 & 4:15:33 & +25:03 \\
B14  & 4:39:59 & +25:44	&  3 & B211 & 4:17:12 & +27:48 \\
B18  & 4:31:13 & +24:21	& 60 & B212 & 4:19:14 & +25:18 \\
B19  & 4:33:00 & +26:16	& 60 & B213 & 4:21:10 & +27:03 \\
B22  & 4:38:00 & +26:03	&120 & B214 & 4:21:55 & +28:32 &  5 \\
B23  & 4:40:33 & +29:52	&  5 & B215 & 4:23:34 & +25:02 \\
B24  & 4:42:53 & +29:44	&  8 & B216 & 4:23:59 & +26:37 \\
B26  & 4:54:38 & +30:37	&  5 & B217 & 4:27:38 & +26:07 \\
B27  & 4:55:08 & +30:33	&  5 & B218 & 4:28:09 & +26:16 & 15 \\
B28  & 4:55:52 & +30:38	&  4 & B219 & 4:34:00 & +29:35 &120 \\
B29  & 5:06:23 & +31:35	& 10 & B220 & 4:41:30 & +25:59 &  7 \\
B207 & 4:04:35 & +26:20	&  & B221 & 4:44:00 & +31:44 & 45 \\
B208 & 4:11:32 & +25:09	&  & B222 & 5:08:23 & +32:10 & 10 \\
\noalign{\smallskip}
\tableline
\end{tabular}
}
\end{center}
\end{table}

Stellar colors from 2MASS clearly demonstrate the two main features
of dark clouds (Figure \ref{fig:colors}). Extinction by dust grains
in the cloud
reddens background stars \citep{tru30,tru34}.  In the optical, the
ratio of total to selective extinction, $R_V = A_V / E_{B-V}$, where
$A_V$ is the visual extinction in magnitudes and $E_{B-V}$ is the
$B-V$ color excess, is a convenient way to compare the physical
properties of dust grains among dark clouds \citep[e.g.,][]
{car89,mat90}. Most diffuse clouds have $R_V \approx$ 3.1; many
dark clouds, including regions in Taurus-Auriga with $A_V >$ 3,
have $R_V \sim$ 3.5--5 \citep{car89,mat90,whit01}.

Figure \ref{fig:colors} also shows that the reddest stars lie
within the dark cloud.
After correcting for extinction, many T Tauri stars have redder
infrared (IR) colors than main sequence stars of similar spectral type
\citep[e.g.,][]{men66,men68,coh73,rss76,ryd82}. These IR excesses
demonstrate that the youngest stars are surrounded by dust grains
with a large range in temperatures \citep{men66,men68,gla74,coh79,
ryd82,mye87,ada87,ada88,ken87,cal94,ken95,and05,furl06}. Images with
{\it HST}
and large ground-based optical/near-IR and radio telescopes reveal
that this dust lies in a circumstellar disk or an infalling envelope
\citep{burrows96,lay97,dut98,duv98,krist98,stap98a,padg99,monin00,
bello02,mcc02,park02,duc03a,sch03,stap03,kri05,kud08}.

With no luminous O or B stars, the Taurus-Auriga dark clouds are mostly
neutral and contain large amounts, 3--4 $\times~10^4~M_{\odot}$, of
molecular gas.  Comprehensive surveys in $^{12}$C$^{16}$O, $^{13}$C$^{16}$O,
and OH confirm the filamentary structure observed in the optical and
near-IR \citep{duv86,ung87,kra91,zho94,abe94,abe95,oni96,oni98,oni02,bli97,
cod97,juv97,gold08,nara08}. Within these structures, the radio observations
reveal dense clumps and cores of molecular gas with masses of
1--100~$M_{\odot}$. The higher density tracers $^{12}$C$^{17}$O,
$^{12}$C$^{18}$O, CS, and NH$_3$ reveal structure within the cores,
including evidence for turbulence on small scales
\citep{ben84,ben89,hay94,jij99}.

\begin{table}
\caption{Lynds' Dark Nebulae in Taurus-Auriga$^1$}
\begin{center}
{\scriptsize
\begin{tabular}[t]{l c c c c c c c c}
\tableline
\noalign{\smallskip}
L & $\alpha$(2000) & $\delta$(2000) & $l$ (deg) & $b$ (deg) & Area (deg$^2$) &
Opac$^2$ & Barnard$^3$ & TGU$^4$ \\
\noalign{\smallskip}
\tableline
\noalign{\smallskip}
1484 & 4:03.1 & 29:08 & 165.72 &$-$17.40 & 0.445 &1& & 1121 \\
1486 & 4:08.1 & 29:07 & 166.55 &$-$16.64 & 0.480 &1& & 1211\\
1489 & 4:04.7 & 26:28 & 167.96 &$-$19.05 & 0.027 &5& & 1144\\
1491 & 4:04.6 & 26:17 & 168.08 &$-$19.20 & 0.004 &5& & 1144\\
1495 & 4:18.1 & 27:37 & 169.27 &$-$16.13 & 2.600 &5&7,10,211,209,216 & 1211\\
1496 & 4:43.2 & 32:05 & 169.58 &$-$09.08 & 0.710 &3& & 1157\\
1497 & 4:27.1 & 28:36 & 169.92 &$-$13.99 & 0.725 &2& & 1158\\
1498 & 4:11.0 & 24:57 & 170.14 &$-$19.11 & 0.118 &5& & 1155\\
1499 & 4:10.5 & 24:47 & 170.18 &$-$19.30 & 0.072 &3& & 1155\\
1500 & 4:33.1 & 29:26 & 170.19 &$-$12.46 & 0.972 &3&219 & 1158\\
1501 & 4:15.0 & 25:07 & 170.68 &$-$18.34 & 1.200 &2& & 1180\\
1503 & 4:40.4 & 29:55 & 170.85 &$-$10.95 & 0.008 &5&23&1169\\
1504 & 4:41.2 & 29:55 & 170.97 &$-$10.82 & 0.336 &1&&1169\\
1506 & 4:20.0 & 25:17 & 171.37 &$-$17.40 & 0.334 &6&&1180\\
1507 & 4:43.2 & 29:45 & 171.38 &$-$10.60 & 0.090 &5&24&1169\\
1508 & 4:41.1 & 29:05 & 171.62 &$-$11.36 & 0.183 &3&&1169\\
1511 & 4:20.0 & 24:47 & 171.75 &$-$17.74 & 0.300 &3&&1173\\
1513 & 4:52.2 & 30:49 & 171.79 &$-$08.41 & 0.057 &4&&1187\\
1514 & 4:42.6 & 29:05 & 171.83 &$-$11.11 & 0.050 &4&&1169\\
1515 & 4:53.2 & 30:54 & 171.87 &$-$08.19 & 0.147 &3&&1187\\
1517 & 4:55.2 & 30:34 & 172.40 &$-$08.06 & 0.051 &6&28,27,26&1187\\
1519 & 4:55.7 & 30:34 & 172.47 &$-$07.98 & 0.063 &4&&1187\\
1520 & 4:44.1 & 28:25 & 172.57 &$-$11.29 & 0.398 &1&&1190\\
1521 & 4:33.1 & 26:06 & 172.76 &$-$14.66 & 4.100 &4&22,19,14,220&1211\\
1522 & 5:07.2 & 32:03 & 172.79 &$-$05.11 & 0.018 &4&222&1193\\
1523 & 5:06.2 & 31:43 & 172.92 &$-$05.48 & 0.017 &6&29&1193\\
1524 & 4:28.0 & 24:36 & 173.16 &$-$16.50 & 0.324 &5&215,212,210&1198\\
1527 & 4:39.1 & 26:15 & 173.53 &$-$13.53 & 0.010 &6&&1211\\
1528 & 4:37.1 & 25:46 & 173.63 &$-$14.20 & 2.000 &3&&1211\\
1529 & 4:32.0 & 24:26 & 173.91 &$-$15.92 & 0.223 &5&18&1198\\
1531 & 4:32.0 & 24:19 & 174.00 &$-$16.00 & 0.014 &3&&1198\\
1532 & 4:40.1 & 25:45 & 174.08 &$-$13.68 & 0.112 &4&&1211\\
1533 & 4:36.2 & 24:55 & 174.18 &$-$14.88 & 0.011 &5&&1210\\
1534 & 4:40.1 & 25:35 & 174.21 &$-$13.79 & 0.870 &5&&1211\\
1535 & 4:35.5 & 23:54 & 174.87 &$-$15.66 & 0.111 &6&&1198\\
1536 & 4:33.0 & 23:06 & 175.12 &$-$16.62 & 1.500 &4&&1227\\
1537 & 4:51.1 & 26:25 & 175.15 &$-$11.34 & 0.382 &3&&1211\\
1538 & 4:46.1 & 25:05 & 175.49 &$-$13.06 & 3.000 &4&&1211\\
1539 & 4:56.1 & 26:34 & 175.72 &$-$10.36 & 2.230 &1&&1228\\
1540 & 5:01.1 & 26:09 & 176.75 &$-$09.73 & 0.282 &4&&1247\\
1541 & 4:44.0 & 22:45 & 177.07 &$-$14.88 & 0.352 &1&&$\ldots$\\
1542 & 5:01.1 & 25:34 & 177.22 &$-$10.08 & 0.626 &3&&1247\\
1543 & 4:27.4 & 18:51 & 177.66 &$-$20.35 & 0.090 &5&&1246\\
1544 & 5:04.1 & 25:14 & 177.91 &$-$09.74 & 0.109 &6&&1247\\
1545 & 5:15.3 & 26:43 & 178.18 &$-$06.82 & 0.052 &3&&1253\\
1546 & 4:28.9 & 18:26 & 178.25 &$-$20.34 & 0.370 &4&&1246\\
1547 & 5:16.1 & 26:23 & 178.55 &$-$06.86 & 0.049 &5&&1253\\
1548 & 5:16.1 & 26:13 & 178.69 &$-$06.96 & 0.061 &4&&1253\\
1549 & 5:17.1 & 26:13 & 178.82 &$-$06.78 & 0.079 &4&&1253\\
1551 & 4:31.4 & 18:06 & 178.92 &$-$20.10 & 0.043 &6&&1246\\
1556 & 4:37.4 & 16:55 & 180.85 &$-$19.71 & 0.050 &5&&1277\\
1558 & 4:45.9 & 17:05 & 182.02 &$-$18.00 & 0.713 &2&&1295\\
\noalign{\smallskip}
\tableline
\noalign{\smallskip}

\multicolumn{9}{l}{\parbox{0.95\textwidth}{\footnotesize
    $^1$ This table is based on an updated version of the published
    catalog downloaded from the CDS.}}\\[1ex]

\multicolumn{9}{l}{\parbox{0.8\textwidth}{\footnotesize
    $^2$ Opacity class from Lynds (1962)}}\\[1ex]

\multicolumn{9}{l}{\parbox{0.8\textwidth}{\footnotesize
    $^3$ Barnard dark nebulae from Table 1.}}\\[1ex]

\multicolumn{9}{l}{\parbox{0.8\textwidth}{\footnotesize
    $^4$ Dobashi et al. (2005) high resolution identification from
    the association file named `hassoc.dat' and visual examination of
    Figure 18--5--6, which plots Lynds' IDs on the extinction maps.}}

\end{tabular}
}
\end{center}
\end{table}

Multiwavelength observations probe structure on the smallest scales within
the Taurus-Auriga clouds. High angular resolution radio observations of
molecular cloud cores detect infalling material and show that this gas is
distinct from lower density gas in the molecular outflow
\citep{oha91,bar93b,oha96,oha97a,oha97b,cha00,hog00,hog01}.
In some cases, radio observations detect the orbital rotation of
material in the disk surrounding the protostar \citep[e.g.,][]
{sar91,kaw93,ter93,hay93,gui94,koe95,mun96,wil96,andre99,bello02}.
In cores with young protostars, near-IR and optical imaging and
polarimetric imaging reveal the structure of the disk and inner
envelope, along with beautiful, often bipolar, reflection nebulae
at the interface between the infalling gas and the bipolar outflow
\citep{tam91,ken93b,luc96,luc97,luc98,wood96,wood98,wood01,wood02,
whi97,har99,padg99,cot01}.

Radio observations also probe the densest parts of the Taurus-Auriga clouds.
In TMC-1/L1534, for example, high densities provide a good laboratory for the
production of complex organic and inorganic molecules.  In the last 25 years,
detections of
C$_3$O \citep{bro85}, HCN \citep{irv84}, HC$_{3}$N \citep{irv84},
HC$_{5}$N \citep{takano90}, HC$_{11}$N \citep{bel85}, HCCNC \citep{kaw92},
CH$_{2}$CN \citep{irv88}, CH$_{3}$CN \citep{minh93},
H$_2$C$_6$ \citep{lan97}, C$_3$H$_2$ \citep{fos01},
HDCS \citep{min97,kai04}, C$_8$H$^{-}$ \citep{bru07}, and
other molecules show that cloud cores have a rich chemistry
prior to the formation of circumstellar disks and planets.

Ices are an important part of dark cloud chemistry. Measurements of the
absorption along the line-of-sight to field stars behind the dark cloud
yields measures of ice abundances and an insight into the
formation and evolution of grains and complex molecules within
the cloud \citep{boo02,vanD04,zas07}. In Taurus-Auriga, detections of
H$_2$O \citep{whit88,bar90,smi93} and CO$_2$ \citep{whit01,ber05}
show that water ice is the dominant constituent, with CO$_2$
roughly 25\% as abundant as water ice.  These observations
demonstrate
that there is an extinction threshold for ice absorption within
dark clouds. The threshold varies from cloud to cloud, with
$A_V \sim$ 3 mag in Taurus-Auriga \citep{whit88} and $A_V \sim$ 12 mag
in Ophiuchus \citep{tan90}. Some observations show that the ice
absorption varies along the line-of-sight to stars within the cloud,
implying that the chemistry close to pre-main sequence stars is
different than along more quiescent sightlines \citep{lei01}.

\begin{figure}[!ht]
\centering
\plotone{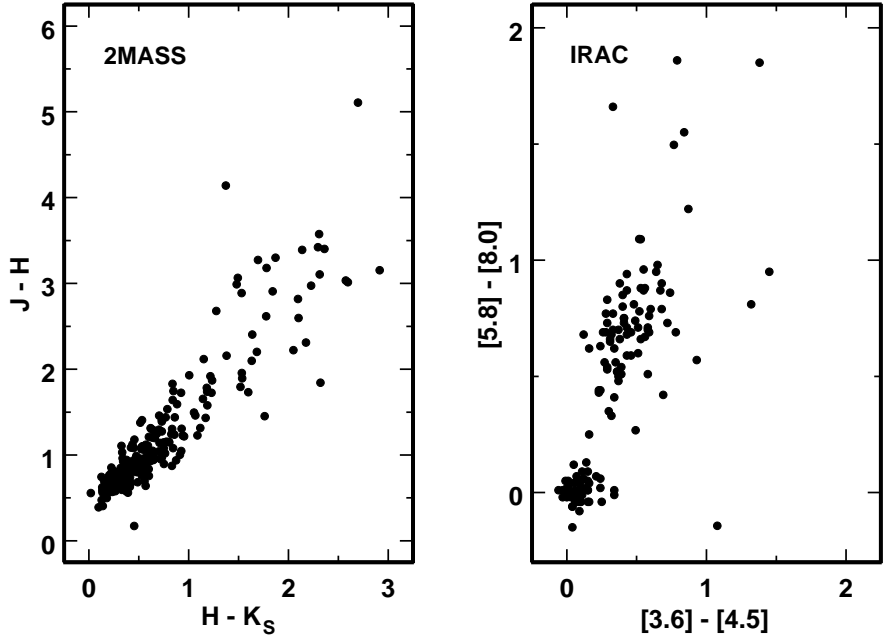}
\vskip -7ex
\caption{Infrared colors for Taurus-Auriga pre-main sequence stars.
Left panel: data from the 2MASS all-sky survey;
right panel: data from the IRAC camera on {\it Spitzer}
\citep{luh06b}.
Weak emission T Tauri stars have $H-K_s < 0.4$ and lie
in a clump at $[3.6]-[4.5]$ $\approx$ $[5.8]-[8.0]$
$\approx$ 0 in the IRAC color-color plot.
Classical T Tauri stars have $H-K_s <$ 1--1.5 and
fall in a band with $[3.6]-[4.5]$ $\approx$ 0.5--1.0
and $[5.8]-[8.0]$ $\approx$ 0.25--0.75.  Embedded
protostars have $ H-K_s >$ 1.25--1.5,
$[3.6]-[4.5]$ $\ge$ 1 and $[5.8]-[8.0]$ $\ge$ 1.
}
\label{fig:ircolors}
\end{figure}

\section{Pre-Main Sequence Population}

The Taurus-Auriga clouds have a rich population of pre-main
sequence stars. Most T Tauri stars discovered before the 1980s
are bright optical variables with G, K, or M spectral types
and strong H~I and Ca~II emission lines \citep{joy45,joy49}.
In the last two decades,
X-ray \citep{walt87,wal88,neu95,gud07a,gud07b,gud07c,sce07},
optical \citep{herb86,gomez92,gomez93,bri93,bri97,gud07a,luh06a,sles06},
near-IR \citep{gomez94,luh98,luh00,luh06a,luh06b,gud07a},
far-IR \citep{bei86,harr88,ken90b,bei92,gud07a},
and radio \citep{bei85}
surveys revealed three populations of young stars
\citep[Figure \ref{fig:ircolors};][]{lad87,ada87,ada88}.
Embedded sources, sometimes known as protostars, are optically
invisible young stars with spectral energy distributions (SEDs)
that peak at mid-IR to far-IR wavelengths.  Classical T Tauri
stars have strong emission lines and substantial IR or UV excesses.
Weak emission T Tauri stars have weak or no emission
lines and negligible excesses.
These objects form a rough age sequence with protostars
as the youngest stars and weak emission T Tauri stars
as the oldest.
Table 3 lists the current sample of young stars and brown dwarfs
in Taurus-Auriga, using data for coordinates and photometry from
the 2MASS all-sky survey \citep{nik00,skr06}, the USNO 1B atlas
\citep{mon03}, and data from
\citet{coh79}, \citet[][1992]{bei86},
\citet{ada90}, \citet{bec90},
\citet[][1993a, 1993b, 1994b, 1998]{ken90b},
\citet[][1993, 1994]{gomez92},
\citet[][1997, 1998, 1999, 2002]{bri93},
\citet{ken95}, \citet{neu95},\citet{wic96},
\citet{luh98}, \citet{ber99}, \citet[2004]{luh00},
\citet[][2006]{luh03}, and
\citet[][2006]{gui05}.

\begin{figure}[p]
\plotone{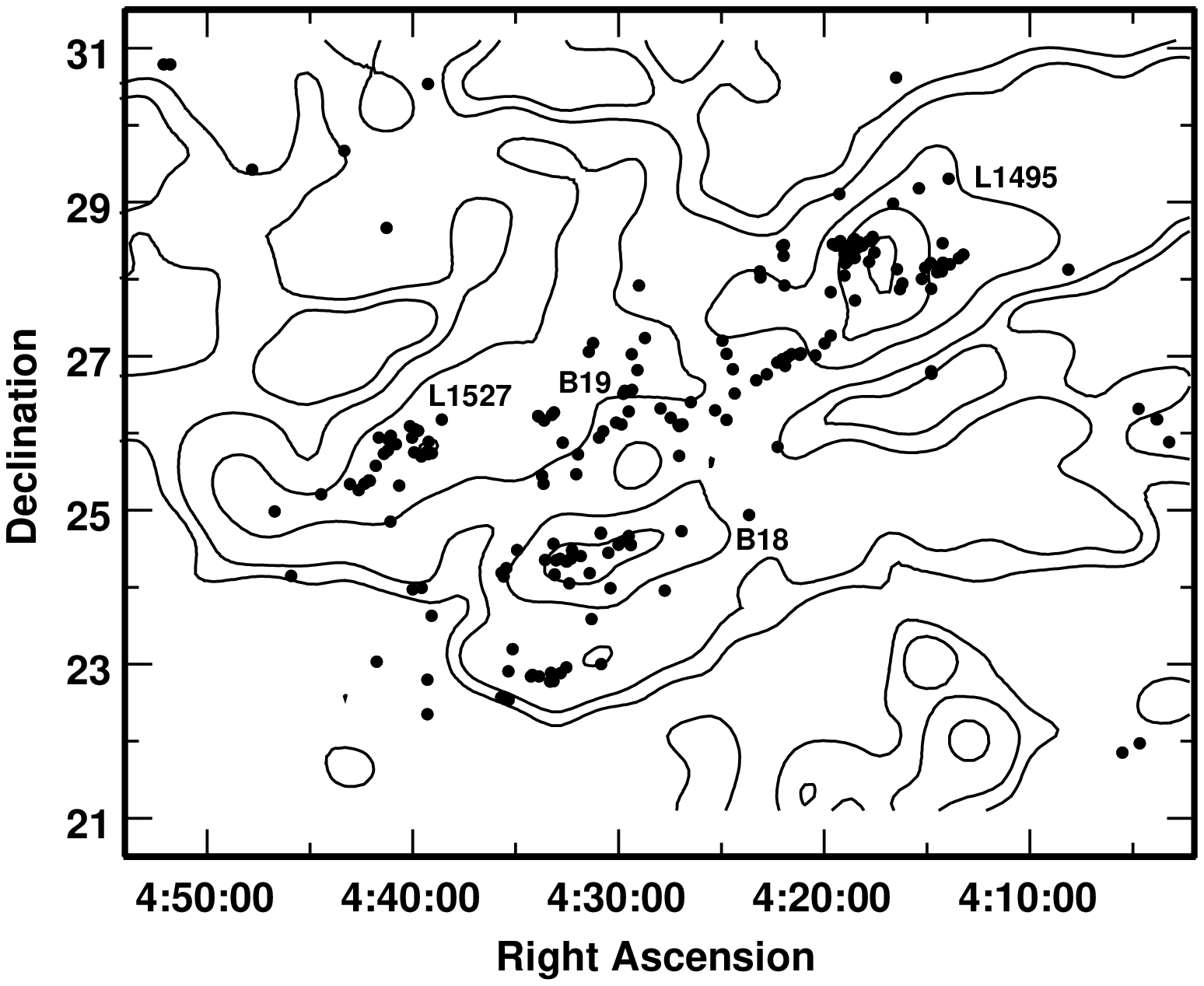}
\vspace{-1mm}
\caption{Sky map for the center of the Taurus-Auriga region in
J2000 coordinates.  Solid contours indicate CO column densities
from \citet{ung87}; the levels are 3, 5, 10, 15, and 20 K km s$^{-1}$.
Solid points indicate the positions of pre-main sequence stars
from Table 3. Groups of young stars lie in
L1495 (NW; RA = $\rm 4^h~12^m$--$\rm 4^h~20^m$, Dec = 27\deg--29\deg),
B18/L1529 (center; RA = $\rm 4^h~24^m$--$\rm 4^h~36$, Dec = 23\deg--25\deg),
B19/L1521 (center; RA = $\rm 4^h~24^m$--$\rm 4^h~36$, Dec = 25\deg--27\deg), and
L1527-29, L1534-35 (E; RA = $\rm 4^h~36^m$--$\rm 4^h~44$, Dec = 24\deg--26\deg).
Only a few young stars lie outside the densest molecular gas.}
\label{fig:co1}

\vspace{5mm}
\includegraphics[width=0.9\textwidth]{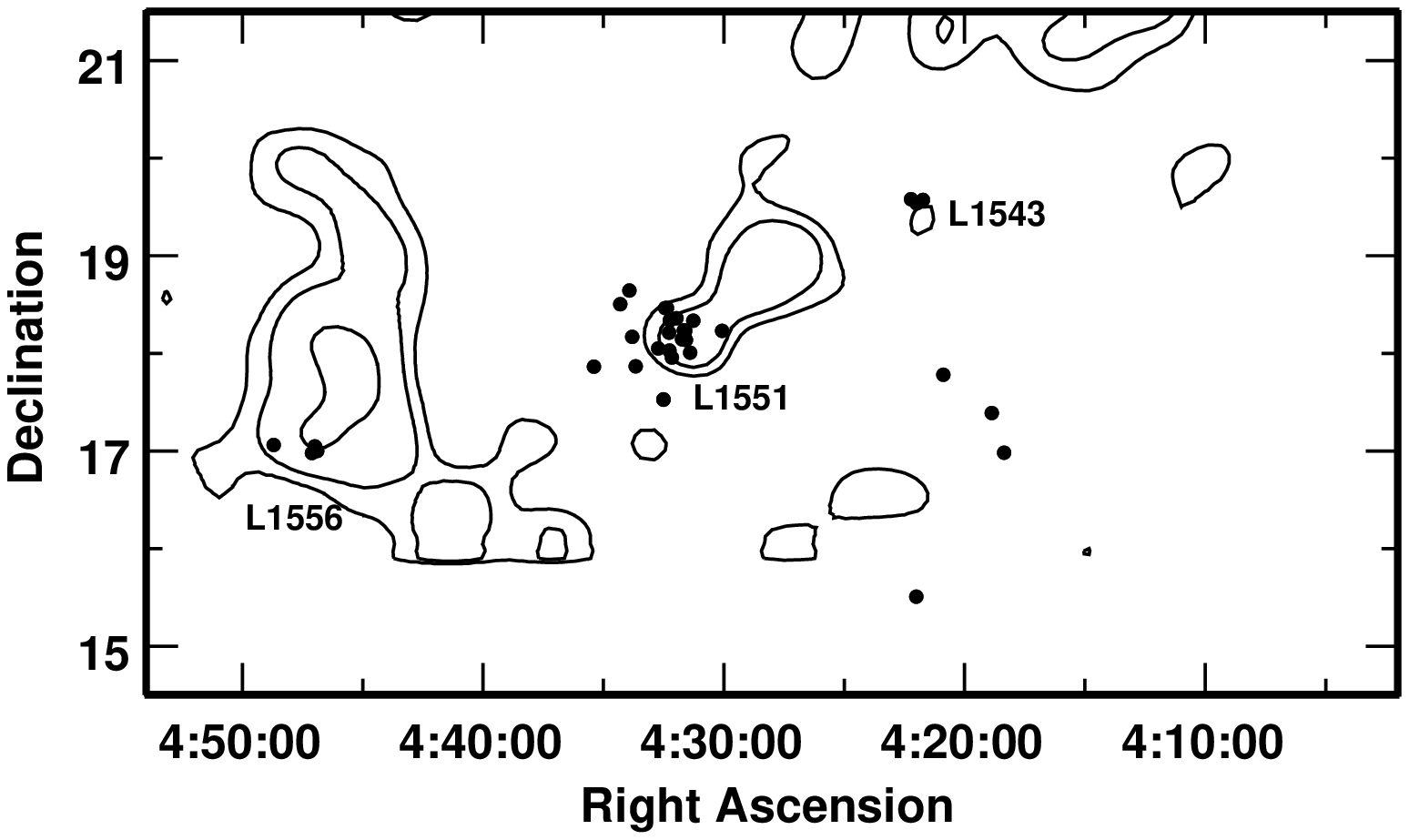}
\vspace{-2mm}
\caption{As in Figure \ref{fig:co1} for the southern portion of the
Taurus-Auriga region.  Groups of young stars are heavily concentrated
in the L1551 dark cloud (RA = $\rm 4^h~31^m$, Dec = 18\deg), with a few
stars in L1543 (RA = $\rm 4^h~23^m$, Dec = 19\deg) and
L1556 (RA = $\rm 4^h~46^m$, Dec = 17\deg).}
\label{fig:co2}
\end{figure}

The surface density of pre-main sequence stars closely follows
the contours of the dark clouds. In the central portion of
Taurus-Auriga (Figure \ref{fig:co1}), most pre-main sequence
stars lie within
the darkest clouds, B7/L1495 to the NW, B18, and B22. Other stars
are projected along the narrow dark lanes that connect the larger
clouds. A few stars are sprinkled at random across the cloud.
Because they are often older, weak emission T Tauri stars are less
concentrated in the dark cloud than classical T Tauri stars.
Embedded protostars are most closely associated with cloud material.

\begin{landscape}
\begin{center}
{ \footnotesize
\begin{longtable}{ l c c r r r | l c c r r r}

\caption{Pre-main sequence stars in  Taurus- Auriga\label{table3a}}\\
\tableline
\noalign{\smallskip}

PMS & $\alpha$(2000) & $\delta$(2000) & \multicolumn{1}{c}{ B} & \multicolumn{1}{c}{R} & \multicolumn{1}{c}{K} & PMS & $\alpha$(2000) & $\delta$(2000) & \multicolumn{1}{c}{ B} & \multicolumn{1}{c}{R} & \multicolumn{1}{c}{K} \\
\noalign{\smallskip}
\tableline
\noalign{\smallskip}
\endfirsthead

\caption{Pre-main sequence stars in  Taurus- Auriga (continued)}\\
\tableline
\noalign{\smallskip}
PMS & $\alpha$(2000) & $\delta$(2000) & \multicolumn{1}{c}{ B} & \multicolumn{1}{c}{R} & \multicolumn{1}{c}{K} & PMS & $\alpha$(2000) & $\delta$(2000) & \multicolumn{1}{c}{ B} & \multicolumn{1}{c}{R} & \multicolumn{1}{c}{K} \\
\noalign{\smallskip}
\tableline
\noalign{\smallskip}
\endhead

\noalign{\smallskip}
\tableline
\endfoot

\noalign{\smallskip}
\tableline
\noalign{\smallskip}
\multicolumn{12}{l}{\parbox{0.8\textwidth}{\footnotesize
    $^1$HP  Tau and HP  Tau/G3 are confused with reflection nebulosity
    on the USNO plates.}}\\[2ex]
\endlastfoot

HBC 351  &  3:52:02.23 & 24:39:47.9 & 12.57 & 10.83 &   9.07 &        J04152409+2910434& 4:15:24.10 & 29:10:43.5 &    & 18.95 & 12.36 \\
HBC 352  &  3:54:29.50 & 32:03:01.3 & 10.56 & 10.70 &  9.58 &        J04153916+2818586& 4:15:39.16 & 28:18:58.6 & 17.58 & 14.77 &  9.24 \\
HBC 353  &  3:54:30.17 & 32:03:04.3 &    &    & 9.86 &        J04155799+2746175& 4:15:57.99 & 27:46:17.5 & 18.60 & 16.80 & 10.52 \\
HBC 354  &  3:54:35.56 & 25:37:11.1 & 12.79 & 11.23 &  11.10 &        J04161210+2756385& 4:16:12.10 & 27:56:38.5 & 19.89 & 17.65 &  10.34 \\
HBC 355  &  3:54:35.97 & 25:37:08.1 & 12.79 & 11.23 &  10.21 &        J04161885+2752155& 4:16:18.86 & 27:52:15.5 & 20.58 & 18.64 & 11.35 \\
HBC 356  &  4:03:13.95 & 25:52:59.7 & 13.95 & 12.35 &  10.16 &        LkCa 4  &  4:16:28.10 & 28:07:35.8 & 14.49 & 11.71 &   8.32 \\
HBC 357  &  4:03:13.95 & 25:52:59.7 & 13.95 & 12.35 &  10.16 &        J04163048+3037053& 4:16:30.48 & 30:37:05.3 & 19.88 & 18.06 &  12.62 \\
HBC 358  &  4:03:49.30 & 26:10:52.0 & 15.57 & 13.50 &   9.46 &        J04163911+2858491& 4:16:39.12 & 28:58:49.1 & 20.35 & 18.73 & 11.28 \\
XEST06-006 &  4:03:49.97 & 26:20:38.2 & 20.36 & 17.83 &  12.34 &        CY Tau  &  4:17:33.72 & 28:20:46.8 & 15.15 & 12.50 &   8.60 \\
HBC 359  &  4:03:50.84 & 26:10:53.1 & 15.10 & 13.12 &   9.53 &        LkCa 5  &  4:17:38.93 & 28:33:00.5 & 15.59 & 12.70 &   9.05 \\
HBC 360  &  4:04:39.36 & 21:58:18.6 & 13.97 & 12.46 &   9.97 &        KPNO-10  &  4:17:49.55 & 28:13:31.8 & 18.36 & 16.14 &  10.79 \\
HBC 361  &  4:04:39.84 & 21:58:21.5 &  0.00 & 12.70 &  10.10 &        V410 X-ray 1 &  4:17:49.65 & 28:29:36.2 & 17.26 & 14.22 &   9.08 \\
IRAS04016+2610 &  4:04:43.22 & 26:18:54.5 & 20.35 & 16.50 &   9.84 &        V410 X-ray 3 &  4:18:07.96 & 28:26:03.6 & 19.62 & 16.92 &  10.45 \\
HBC 362  &  4:05:30.87 & 21:51:10.6 & 14.73 & 13.43 &  10.06 &        J04181078+2519574& 4:18:10.78 & 25:19:57.4 & 15.92 & 12.26 &  9.03 \\
J04080782+2807280& 4:08:07.82 & 28:07:28.1 & 18.32 & 14.97 & 11.39 &        V410  Anon 13 &  4:18:17.10 & 28:28:41.9 &  0.00 & 18.95 &  10.96 \\
J04124858+2749563& 4:12:48.58 & 27:49:56.3 & 17.21 & 15.97 &  11.68 &        HBC 372  &  4:18:21.47 & 16:58:47.0 & 14.32 & 13.50 &  10.46 \\
LkCa 1  &  4:13:14.14 & 28:19:10.8 & 15.74 & 12.84 &   8.62 &        V410  Anon 24 &  4:18:22.39 & 28:24:37.5 &    &    & 10.73 \\
 Anon1  &  4:13:27.22 & 28:16:24.7 & 15.03 & 12.06 &   7.79 &        V410  Anon 25 &  4:18:29.09 & 28:26:19.1 &    &    & 9.94 \\
IRAS04108+2803 A &  4:13:53.28 & 28:11:23.3 &    &    & 10.37 &        KPNO-11  &  4:18:30.30 & 27:43:20.8 & 18.52 & 16.30 &  11.01 \\
IRAS04108+2803 B &  4:13:54.71 & 28:11:32.8 &    &    & 11.06 &        V410 Tau A &  4:18:31.10 & 28:27:16.2 & 11.48 &  9.88 &   7.63 \\
IRAS04108+2910 &  4:13:57.37 & 29:18:19.3 & 17.44 & 14.43 &  9.36&        V410 Tau B &  4:18:31.10 & 28:27:16.2 & 11.48 &  9.88 &   7.63 \\
J04141188+2811535& 4:14:11.88 & 28:11:53.5 & 18.47 & 17.36 & 11.64 &        V410 Tau C &  4:18:31.10 & 28:27:16.2 & 11.48 &  9.88 &   7.63 \\
V773 Tau A &  4:14:12.91 & 28:12:12.4 & 11.39 &  9.97 &   6.21 &        DD Tau A  &  4:18:31.12 & 28:16:29.0 & 15.67 & 13.40 &   7.88 \\
V773 Tau B &  4:14:12.91 & 28:12:12.4 & 11.39 &  9.97 &   6.21 &        DD Tau B  &  4:18:31.12 & 28:16:29.0 & 15.67 & 13.40 &   7.88 \\
FM Tau  &  4:14:13.58 & 28:12:49.2 & 15.57 & 13.14 &   8.76 &        CZ Tau A  &  4:18:31.58 & 28:16:58.5 & 16.27 & 14.50 &   9.36 \\
FN Tau  &  4:14:14.58 & 28:27:58.0 & 16.30 & 13.48 &   8.19 &        CZ Tau B  &  4:18:31.58 & 28:16:58.5 & 16.27 & 14.50 &   9.36 \\
CW Tau  &  4:14:17.00 & 28:10:57.8 & 14.84 & 11.75 &   7.13 &        IRAS04154+2823 &  4:18:32.03 & 28:31:15.3 &    &    & 10.27 \\
CIDA-1  &  4:14:17.60 & 28:06:09.6 & 18.15 & 15.86 &   9.88 &        V410 X-ray 2 &  4:18:34.44 & 28:30:30.2 &    &    & 9.22 \\
IRAS04113+2758 &  4:14:26.26 & 28:06:03.2 & 21.36 & 18.91 &   7.78 &        V410 X-ray 4 &  4:18:40.23 & 28:24:24.5 &    &    & 9.69 \\
MHO2  &  4:14:26.40 & 28:05:59.7 & 21.36 & 18.91 &   7.80 &        V892 Tau  &  4:18:40.61 & 28:19:15.5 & 16.60 & 13.96 &   5.79 \\
MHO3  &  4:14:30.55 & 28:05:14.7 & 20.24 & 16.84 &   8.24 &        LR1  &  4:18:41.33 & 28:27:25.0 &    &    & 11.05 \\
FP Tau  &  4:14:47.30 & 26:46:26.4 & 14.81 & 12.13 &   8.87 &        V410 X-ray  7 &  4:18:42.50 & 28:18:49.8 &  0.00 & 18.49 &  9.26 \\
XEST20-066 &  4:14:47.39 & 28:03:05.5 & 17.33 & 15.09 &   9.92 &        V410  Anon 20 &  4:18:45.05 & 28:20:52.8 &    &    & 11.93 \\
CX Tau  &  4:14:47.86 & 26:48:11.0 & 15.18 & 12.63 &   8.81 &        Hubble4  &  4:18:47.03 & 28:20:07.3 & 13.34 & 11.93 &   7.29 \\
LkCa 3 A  &  4:14:47.97 & 27:52:34.6 & 13.98 & 11.35 &   7.42 &        KPNO-2  &  4:18:51.15 & 28:14:33.2 &  0.00 & 18.95 &  12.75 \\
LkCa 3 B  &  4:14:47.97 & 27:52:34.6 & 13.98 & 11.35 &   7.42 &        CoKu Tau/1 &  4:18:51.47 & 28:20:26.4 & 19.56 & 16.06 &  10.97 \\
FO Tau A  &  4:14:49.28 & 28:12:30.5 & 16.97 & 13.79 &   8.12 &        HBC 376  &  4:18:51.70 & 17:23:16.5 & 12.91 & 11.41 &   9.27 \\
FO Tau B  &  4:14:49.28 & 28:12:30.5 & 16.97 & 13.79 &   8.12 &        IRAS04158+2805 &  4:18:58.13 & 28:12:23.4 & 20.02 & 17.94 &  11.18 \\
CIDA-2  &  4:15:05.15 & 28:08:46.2 & 17.12 & 14.52 &   9.09 &        V410 X-ray  6 &  4:19:01.10 & 28:19:42.0 & 18.95 & 16.50 &  9.13 \\
KPNO-1  &  4:15:14.71 & 28:00:09.6 &    &    & 13.77 &        KPNO-12  &  4:19:01.26 & 28:02:48.7 &    &    & 14.93 \\
V410 X-ray 5a &  4:19:01.97 & 28:22:33.2 & 21.28 & 18.48 &  10.15 &    XEST11-087 &  4:22:24.04 & 26:46:25.8 & 18.71 & 15.91 & 9.77 \\
FQ Tau A  &  4:19:12.81 & 28:29:33.0 & 16.81 & 14.33 &   9.31 &    IRAS04196+2638 &  4:22:47.86 & 26:45:53.0 & 19.56 & 16.10 & 9.29 \\
FQ Tau B  &  4:19:12.81 & 28:29:33.0 & 16.81 & 14.33 &   9.31 &    J04230607+2801194& 4:23:06.07 & 28:01:19.4 & 20.20 & 16.98 &  11.20 \\
 BP Tau  &  4:19:15.83 & 29:06:26.9 & 13.00 & 13.38 &   7.74 &    IRAS04200+2759 &  4:23:07.76 & 28:05:57.3 & 17.71 & 16.66 &  10.41 \\
V819 Tau  &  4:19:26.25 & 28:26:14.2 & 15.26 & 12.01 &   8.42 &    J04231822+2641156& 4:23:18.22 & 26:41:15.6 & 19.58 & 19.34 & 10.18 \\
FR  Tau  &  4:19:35.45 & 28:27:21.8 & 17.26 & 14.84 &   9.97 &    FU Tau  &  4:23:35.39 & 25:03:02.6 & 19.10 & 16.70 &  9.32  \\
LkCa 7 A  &  4:19:41.27 & 27:49:48.4 & 14.23 & 11.76 &   8.26 &    FT Tau  &  4:23:39.19 & 24:56:14.1 & 15.72 & 12.65 &   8.60 \\
LkCa 7 B  &  4:19:41.27 & 27:49:48.4 & 14.23 & 11.76 &   8.26 &    J04242090+2630511& 4:24:20.90 & 26:30:51.1 & 20.34 & 18.73 & 12.43 \\
IRAS04166+2706 &  4:19:41.48 & 27:16:07.0 &    &    & 11.54 &    CFHT-9  &  4:24:26.46 & 26:49:50.3 & 20.59 & 18.30 & 11.76 \\
IRAS04169+2702 &  4:19:58.44 & 27:09:57.0 &    &    & 11.58 &    IRAS04216+2603 &  4:24:44.57 & 26:10:14.1 & 18.33 & 15.46 &   9.05 \\
J04201611+2821325& 4:20:16.11 & 28:21:32.5 & 20.56 & 18.42 & 12.55 &    J1-4423  &  4:24:45.06 & 27:01:44.7 & 17.62 & 15.16 &  10.46 \\
J04202555+2700355& 4:20:25.55 & 27:00:35.5 & 20.55 & 18.75 &  11.51 &    IP Tau  &  4:24:57.08 & 27:11:56.5 & 14.11 & 12.29 &   8.35 \\
J04202583+2819237& 4:20:25.83 & 28:19:23.7 & 20.40 & 18.02 & 11.72 &    J1-4872 A &  4:25:17.67 & 26:17:50.4 & 15.13 & 12.42 &   8.54 \\
J04202606+2804089& 4:20:26.06 & 28:04:08.9 & 16.41 & 13.90 &  9.70 &    J1-4872 B &  4:25:17.67 & 26:17:50.4 & 15.13 & 12.42 &   8.54 \\
XEST16-045 &  4:20:39.18 & 27:17:31.7 & 18.15 & 16.22 &  9.56 &    KPNO-3  &  4:26:29.39 & 26:24:13.7 & 19.89 & 19.02 & 12.08 \\
J2-157  &  4:20:52.73 & 17:46:41.5 & 17.12 & 15.52 &  10.78 &    FV Tau A  &  4:26:53.52 & 26:06:54.3 & 17.47 & 14.00 &   7.44 \\
CFHT-19  &  4:21:07.95 & 27:02:20.4 &  0.00 & 18.85 &  10.54 &    FV Tau B  &  4:26:53.52 & 26:06:54.3 & 17.47 & 14.00 &   7.44 \\
J04210934+2750368& 4:21:09.34 & 27:50:36.8 & 17.77 & 15.55 & 10.36 &    FV Tau/c A &  4:26:54.40 & 26:06:51.0 & 18.69 & 15.46 &   8.87 \\
IRAS04181+2654 B &  4:21:10.38 & 27:01:37.2 &    &    & 11.09 &    FV Tau/c B &  4:26:54.40 & 26:06:51.0 & 18.69 & 15.46 &   8.87 \\
IRAS04181+2655 &  4:21:10.90 & 27:02:06.0 &    &    & &     IRAS04239+2436 &  4:26:56.29 & 24:43:35.3 &    &    &  9.99 \\
IRAS04181+2654 A &  4:21:11.46 & 27:01:09.4 &    &    & 10.34 &    KPNO-13  &  4:26:57.33 & 26:06:28.4 & 19.78 & 16.64 &   9.58 \\
J04213459+2701388& 4:21:34.59 & 27:01:38.8 & 20.46 & 17.32 &  10.44 &    DG  Tau B &  4:27:02.66 & 26:05:30.4 &    & 19.39 & \\
XEST21-026 &  4:21:40.13 & 28:14:22.4 & 18.15 & 16.22 &  11.03 &    DF Tau A  &  4:27:02.80 & 25:42:22.3 & 12.96 & 13.34 &   6.73 \\
IRAS04187+1927 &  4:21:43.23 & 19:34:13.3 & 16.99 & 14.86 &   8.02 &    DF Tau B  &  4:27:02.80 & 25:42:22.3 & 12.96 & 13.34 &   6.73 \\
CFHT-10  &  4:21:46.31 & 26:59:29.6 &    &    & 12.31 &    DG Tau  &  4:27:04.69 & 26:06:16.3 & 10.13 &  8.97 &   6.99 \\
J04215450+2652315& 4:21:54.51 & 26:52:31.5 &    &    & 13.90 &    KPNO-4  &  4:27:27.99 & 26:12:05.2 &    & 19.70 &  13.28 \\
DE Tau  &  4:21:55.63 & 27:55:06.0 & 14.98 & 11.89 &   7.80 &    CFHT-15  &  4:27:45.38 & 23:57:24.3 &    &    &  13.69 \\
RY Tau  &  4:21:57.40 & 28:26:35.5 & 10.73 &  9.62 &   5.39 &    IRAS04248+2612 &  4:27:57.30 & 26:19:18.3 & 20.58 & 15.73 &  11.03 \\
HD283572 &  4:21:58.84 & 28:18:06.6 &  9.44 &  8.56 &   6.87 &    J04284263+2714039& 4:28:42.63 & 27:14:03.9 & 17.51 & 16.81 &  10.46 \\
T TauN  &  4:21:59.43 & 19:32:06.3 & 10.47 &  9.19 &   5.33 &    J04290068+2755033& 4:29:00.68 & 27:55:03.3 &    &    &  12.85 \\
T TauS  &  4:21:59.43 & 19:32:06.3 & 10.47 &  9.19 &   5.33 &    IRAS04260+2642 &  4:29:04.98 & 26:49:07.3 & 19.97 & 18.46 &  11.88 \\
IRAS04191+1523 &  4:22:00.10 & 15:30:21.3 &    &    & 12.26 &    J1-507  &  4:29:20.71 & 26:33:40.6 & 16.08 & 13.56 &   8.79 \\
H6-5 B  &  4:22:00.69 & 26:57:32.4 &    &    & 11.75 &    IRAS04263+2654 &  4:29:21.65 & 27:01:25.9 &    & 17.96 &  8.72 \\
FS Tau A  &  4:22:02.17 & 26:57:30.4 & 13.78 & 11.33 &   8.18 &    GV Tau A  &  4:29:23.73 & 24:33:00.2 & 17.62 & 13.13 &   8.05 \\
FS Tau B  &  4:22:02.17 & 26:57:30.4 & 13.78 & 11.33 &   8.18 &    GV Tau B  &  4:29:23.73 & 24:33:00.2 & 17.62 & 13.13 &   8.05 \\
LkCa 21  &  4:22:03.13 & 28:25:38.9 & 15.32 & 12.45 &   8.45 &    IRAS04263+2426 &  4:29:24.11 & 24:32:57.7 & 20.54 & 15.36 &  10.57 \\
J04221332+1934392& 4:22:13.32 & 19:34:39.2 & 19.86 & 19.34 & 11.53 &    FW Tau C  &  4:29:29.71 & 26:16:53.2 & 17.47 & 14.65 &   9.39 \\
XEST11-078 &  4:22:15.68 & 26:57:06.0 & 20.18 & 17.16 & 12.03 &    FW Tau A  &  4:29:29.71 & 26:16:53.2 & 17.47 & 14.65 &   9.39 \\
CFHT-14  &  4:22:16.44 & 25:49:11.8 & 20.49 & 19.07 & 11.94 &    FW Tau B  &  4:29:29.71 & 26:16:53.2 & 17.47 & 14.65 &   9.39 \\
CFHT-21  &  4:22:16.76 & 26:54:57.1 & 19.36 & 15.61 & 9.00 &    IRAS04264+2433 &  4:29:30.08 & 24:39:55.0 &    & 18.42 &  11.13 \\
J04293606+2435556& 4:29:36.06 & 24:35:55.6 & 18.32 & 16.17 & 8.66  &   J04320329+2528078& 4:32:03.29 & 25:28:07.8 & 19.71 & 16.84 &  10.72 \\
DH Tau A  &  4:29:41.55 & 26:32:58.2 & 14.96 & 12.09 &   8.18 &   L1551-51 &  4:32:09.27 & 17:57:22.8 & 13.83 & 11.13 &  12.15 \\
DH Tau B  &  4:29:41.55 & 26:32:58.2 & 14.96 & 12.09 &   8.18 &   V827 Tau  &  4:32:14.56 & 18:20:14.7 & 14.35 & 11.39 &   8.23 \\
DI Tau A  &  4:29:42.47 & 26:32:49.3 & 14.12 &  0.00 &   8.39 &   Haro 6-13 &  4:32:15.40 & 24:28:59.7 & 18.86 & 14.85 &   8.10 \\
DI Tau B  &  4:29:42.47 & 26:32:49.3 & 14.12 &  0.00 &   8.39 &   V826 Tau A &  4:32:15.83 & 18:01:38.7 & 14.23 & 11.28 &   8.25 \\
KPNO-5  &  4:29:45.68 & 26:30:46.8 & 20.56 & 18.68 & 11.54 &   V826 Tau B &  4:32:15.83 & 18:01:38.7 & 14.23 & 11.28 &   8.25 \\
IQ Tau  &  4:29:51.56 & 26:06:44.8 & 15.60 & 12.14 &   7.78 &   MHO5  &  4:32:16.02 & 18:12:46.4 & 18.95 & 16.36 &  10.06 \\
CFHT-20  &  4:29:59.50 & 24:33:07.8 & 20.52 & 17.15 &  9.81 &   CFHT-7  &  4:32:17.86 & 24:22:14.9 & 19.99 & 16.73 &  10.38 \\
UX Tau C  &  4:30:03.99 & 18:13:49.3 & 12.04 & 10.30 &   7.55 &   V928 Tau A &  4:32:18.85 & 24:22:27.1 & 15.59 & 12.87 &   8.11 \\
UX Tau B  &  4:30:03.99 & 18:13:49.3 & 12.04 & 10.30 &   7.55 &   V928 Tau B &  4:32:18.85 & 24:22:27.1 & 15.59 & 12.87 &   8.11 \\
UX Tau A  &  4:30:03.99 & 18:13:49.3 & 12.04 & 10.30 &   7.55 &   MHO6  &  4:32:22.10 & 18:27:42.6 & 17.00 & 15.30 &  10.65 \\
KPNO-6  &  4:30:07.24 & 26:08:20.7 &  0.00 & 19.42 &  13.69 &   J04322329+2403013& 4:32:23.29 & 24:03:01.3 & 21.15 & 18.35 &  11.33 \\
CFHT-16  &  4:30:23.65 & 23:59:12.9 & 18.14 & 15.37 &  13.70 &   J04322415+2251083& 4:32:24.15 & 22:51:08.3 & 18.75 & 15.95 &  10.53 \\
FX Tau A  &  4:30:29.61 & 24:26:45.0 & 15.01 & 12.71 &   7.92 &   MHO7  &  4:32:26.27 & 18:27:52.1 & 18.06 & 15.96 &  10.17 \\
FX Tau B  &  4:30:29.61 & 24:26:45.0 & 15.01 & 12.71 &   7.92 &   GG Tau Ba  &  4:32:30.28 & 17:31:30.3 &    &    &  9.97\\
DK Tau A  &  4:30:44.25 & 26:01:24.4 & 14.17 & 11.08 &   7.10 &   GG Tau Bb  &  4:32:30.28 & 17:31:30.3 &    &    &  9.97\\
DK Tau B  &  4:30:44.25 & 26:01:24.4 & 14.17 & 11.08 &   7.10 &   GG Tau Aa  &  4:32:30.34 & 17:31:40.6 & 12.64 & 11.80 &   7.36 \\
IRAS04278+2253 &  4:30:50.28 & 23:00:08.8 & 17.08 & 14.18 &   5.86 &   GG Tau Ab  &  4:32:30.34 & 17:31:40.6 & 12.64 & 11.80 &   7.36 \\
ZZ Tau  &  4:30:51.37 & 24:42:22.2 & 15.82 & 13.31 &   8.44 &   FY Tau  &  4:32:30.58 & 24:19:57.2 & 17.37 & 13.79 &   8.05 \\
ZZ  Tau IRS &  4:30:51.71 & 24:41:47.5 & 18.59 & 16.28 &  10.31 &   FZ Tau  &  4:32:31.76 & 24:20:02.9 & 17.11 & 13.85 &   7.35 \\
KPNO-7  &  4:30:57.18 & 25:56:39.4 &  0.00 & 19.28 &  13.27 &   IRAS04295+2251 &  4:32:32.05 & 22:57:26.6 &  0.00 & 20.18 &  10.14 \\
JH56  &  4:31:14.44 & 27:10:17.9 & 13.77 & 12.20 &   8.79 &   UZ Tau Ba  &  4:32:42.82 & 25:52:31.4 & 13.41 & 11.37 &   7.47 \\
MHO9  &  4:31:15.94 & 18:20:07.2 & 17.66 & 15.16 &  10.30 &   UZ Tau Bb  &  4:32:42.82 & 25:52:31.4 & 13.41 & 11.37 &   7.47 \\
J04311907+2335047& 4:31:19.07 & 23:35:04.7 &    & 19.43 & 12.20 &   UZ Tau A  &  4:32:43.03 & 25:52:31.1 & 13.41 & 11.37 &   7.35 \\
V927 Tau A &  4:31:23.82 & 24:10:52.9 & 16.12 & 13.78 &   8.77 &   L1551-55 &  4:32:43.73 & 18:02:56.3 & 15.29 & 13.31 &   9.31 \\
V927 Tau B &  4:31:23.82 & 24:10:52.9 & 16.12 & 13.78 &   8.77 &   JH112  &  4:32:49.11 & 22:53:02.7 & 16.54 & 13.26 &   8.17 \\
MHO4  &  4:31:24.06 & 18:00:21.5 & 20.31 & 17.50 & &    J04324938+2253082& 4:32:49.38 & 22:53:08.2 & 20.15 &    &   9.20 \\
CFHT-13  &  4:31:26.69 & 27:03:18.8 &    &    &  13.45 &   CFHT-5  &  4:32:50.26 & 24:22:11.5 &    &    &  11.28 \\
L1551 IRS5 &  4:31:34.07 & 18:08:04.9 & 20.11 & 17.10 &   9.82 &   J04325119+1730092& 4:32:51.19 & 17:30:09.2 &    &    &  13.55 \\
LkH$\alpha$ 358  &  4:31:36.13 & 18:13:43.2 & 19.40 & 16.50 &   9.69 &   MHO8  &  4:33:01.97 & 24:21:00.0 & 19.53 & 16.52 &   9.73 \\
HH30 IRS &  4:31:37.50 & 18:12:24.4 & 19.48 & 16.77 &  14.24 &   GH Tau A  &  4:33:06.22 & 24:09:33.9 & 14.57 & 11.63 &   7.79 \\
HL Tau  &  4:31:38.43 & 18:13:57.6 & 14.20 & 10.63 &   7.41 &   GH Tau B  &  4:33:06.22 & 24:09:33.9 & 14.57 & 11.63 &   7.79 \\
XZ Tau A  &  4:31:40.07 & 18:13:57.1 & 16.11 & 13.56 &   7.29 &   V807 Tau A &  4:33:06.64 & 24:09:54.9 & 11.68 & 10.67 &   6.96 \\
XZ Tau B  &  4:31:40.07 & 18:13:57.1 & 16.11 & 13.56 &   7.29 &   V807 Tau B &  4:33:06.64 & 24:09:54.9 & 11.68 & 10.67 &   6.96 \\
L1551NE  &  4:31:44.44 & 18:08:31.5 &    &    &  11.41 &   KPNO-14  &  4:33:07.81 & 26:16:06.6 & 20.60 & 18.70 &  10.27 \\
HK Tau A  &  4:31:50.56 & 24:24:18.0 & 17.18 & 14.40 &   8.59 &   CFHT-12  &  4:33:09.46 & 22:46:48.7 &    & 19.92 &  11.55 \\
HK Tau B  &  4:31:50.56 & 24:24:18.0 & 17.18 & 14.40 &   8.59 &   V830 Tau  &  4:33:10.03 & 24:33:43.3 & 13.33 & 10.95 &   8.42 \\
V710 Tau A &  4:31:57.70 & 18:21:37.0 & 14.95 & 12.70 &   8.69 &   IRAS04301+2608 &  4:33:14.36 & 26:14:23.5 &  0.00 & 19.26 &  12.48 \\
V710 Tau B &  4:31:57.70 & 18:21:37.0 & 14.95 & 12.70 &   8.69 &   IRAS04302+2247 &  4:33:16.50 & 22:53:20.4 &  0.00 & 18.04 &  11.72 \\
J1-665  &  4:31:58.44 & 25:43:29.9 & 17.43 & 14.91 &   9.56 &   IRAS04303+2240 &  4:33:19.07 & 22:46:34.2 & 20.39 & 16.22 &   7.67 \\
XEST17-036 &  4:33:26.21 & 22:45:29.3 & 20.47 & 17.49 &   9.92 &   KPNO-9  &  4:35:51.43 & 22:49:11.9 &    &    & 14.19 \\
GI Tau  &  4:33:34.05 & 24:21:17.0 &    &    &   7.89 &   XEST08-047 &  4:35:52.09 & 22:55:03.9 & 18.56 & 15.85 &  9.81 \\
GK Tau  &  4:33:34.56 & 24:21:05.8 &    &    &   7.47 &   HP Tau  &  4:35:52.77 & 22:54:23.1 &  0.00 &  0.00 &   7.62 \\
IS Tau A  &  4:33:36.78 & 26:09:49.2 & 17.44 & 14.35 &   8.64 &   XEST08-049 &  4:35:52.86 & 22:50:58.5 & 18.50 & 15.63 &   9.75 \\
IS Tau B  &  4:33:36.78 & 26:09:49.2 & 17.44 & 14.35 &   8.64 &   HP Tau/G3 &  4:35:53.49 & 22:54:08.9 &    &   &  8.80 \\
DL Tau  &  4:33:39.06 & 25:20:38.2 & 13.81 & 11.52 &   7.96 &   HP Tau/G2 &  4:35:54.15 & 22:54:13.4 & 11.66 & 10.06 &   7.23 \\
DL Tau  &  4:33:39.06 & 25:20:38.2 & 13.81 & 11.52 &   7.96 &   Haro 6-28 A &  4:35:56.84 & 22:54:36.0 & 16.91 & 15.39 &   9.53 \\
HN Tau B  &  4:33:39.35 & 17:51:52.3 & 15.50 & 13.31 &   8.38 &   Haro 6-28 B &  4:35:56.84 & 22:54:36.0 & 16.91 & 15.39 &   9.53 \\
HN Tau A  &  4:33:39.35 & 17:51:52.3 & 15.50 & 13.31 &   8.38 &   XEST09-042 &  4:35:58.92 & 22:38:35.3 & 13.90 & 11.44 &   8.37 \\
J04333905+2227207& 4:33:39.05 & 22:27:20.7 & 19.08 & 16.13 &  10.71 &   J04361030+2159364& 4:36:10.31 & 21:59:36.5 &    & 20.16 & 13.65 \\
J04334291+2526470& 4:33:42.91 & 25:26:47.0 &    &    &  13.33 &   CFHT-2  &  4:36:10.38 & 22:59:56.0 &    & 19.17 &  13.75 \\
J04334465+2615005& 4:33:44.65 & 26:15:00.5 & 20.23 & 17.91 &   9.74 &   LkCa 14  &  4:36:19.09 & 25:42:58.9 & 12.18 & 10.73 &   8.58 \\
DM Tau  &  4:33:48.71 & 18:10:09.9 & 15.61 & 13.61 &   9.52 &   J04362151p2351165& 4:36:21.51 & 23:51:16.5 & 18.91 & 17.43 & 12.24 \\
CI Tau  &  4:33:52.00 & 22:50:30.1 & 14.30 & 11.85 &   7.79 &   CFHT-3  &  4:36:38.96 & 22:58:11.9 &    & 19.21 &  13.72 \\
XEST17-059 &  4:33:52.52 & 22:56:26.9 & 17.31 & 14.78 &   9.11 &   J04373705+2331080& 4:37:37.05 & 23:31:08.0 & 20.23 & 17.91 & 15.44 \\
J04335245p2612548& 4:33:52.45 & 26:12:54.8 &    &    &  13.99 &   ITG 1  &  4:37:56.70 & 25:46:22.9 & 19.14 & 17.45 &  12.70 \\
IT Tau B  &  4:33:54.70 & 26:13:27.5 & 15.95 & 13.06 &   7.86 &   ITG 2  &  4:38:00.83 & 25:58:57.2 & 20.41 & 18.16 &  10.10 \\
IT Tau A  &  4:33:54.70 & 26:13:27.5 & 15.95 & 13.06 &   7.86 &   J04381486+2611399& 4:38:14.86 & 26:11:39.9 & 19.84 & 19.45 &  12.98 \\
J2-2041  &  4:33:55.46 & 18:38:39.0 & 16.44 & 14.47 &   9.61 &   GM Tau  &  4:38:21.34 & 26:09:13.7 & 18.39 & 16.16 &  10.63 \\
JH108  &  4:34:10.99 & 22:51:44.5 & 16.68 & 13.90 &   9.43 &   DO Tau  &  4:38:28.58 & 26:10:49.4 & 14.61 & 12.12 &   7.30 \\
CFHT-1  &  4:34:15.27 & 22:50:31.0 &    &    & 11.85 &   HV Tau A  &  4:38:35.28 & 26:10:38.6 & 15.02 & 11.88 &   7.91 \\
HBC 407  &  4:34:18.03 & 18:30:06.6 & 14.44 & 13.00 &   9.90 &   HV Tau B  &  4:38:35.28 & 26:10:38.6 & 15.02 & 11.88 &   7.91 \\
XEST08-003 &  4:34:56.93 & 22:58:35.8 & 16.00 & 13.46 &   9.27 &   HV Tau C  &  4:38:35.49 & 26:10:41.5 & 15.02 & 11.88 &   7.91 \\
 A A Tau  &  4:34:55.42 & 24:28:53.1 & 14.39 & 12.02 &   8.05 &   CFHT-6  &  4:39:03.96 & 25:44:26.4 & 19.99 & 18.92 &  11.37 \\
J04345973+2807017& 4:34:59.73 & 28:07:01.7 & 20.15 & 20.45 &  14.65 &   J04390525p2337450& 4:39:05.25 & 23:37:45.0 & 17.30 & 14.55 &  11.55 \\
CFHT-11  &  4:35:08.50 & 23:11:39.8 & 20.35 & 18.01 &  11.59 &   IRAS04361p2547 &  4:39:13.89 & 25:53:20.8 &    &    & 10.72 \\
HO Tau  &  4:35:20.20 & 22:32:14.6 & 15.60 & 14.06 &  10.24 &   CIDA-13  &  4:39:15.86 & 30:32:07.4 & 17.81 & 15.76 &  11.83 \\
FF Tau A  &  4:35:20.89 & 22:54:24.2 & 16.63 & 12.82 &   8.93 &   VY Tau A  &  4:39:17.41 & 22:47:53.3 & 14.70 & 13.03 &   8.96 \\
FF Tau B  &  4:35:20.89 & 22:54:24.2 & 16.63 & 12.82 &   8.93 &   VY Tau B  &  4:39:17.41 & 22:47:53.3 & 14.70 & 13.03 &   8.96 \\
HBC 412 A  &  4:35:24.10 & 17:51:41.0 &    &    & 9.10 &   LkCa 15  &  4:39:17.79 & 22:21:03.4 & 12.54 & 11.43 &   8.16 \\
HBC 412 B  &  4:35:24.10 & 17:51:41.0 &    &    & 9.10 &   GN Tau A  &  4:39:20.90 & 25:45:02.1 & 17.34 & 13.76 &  8.06\\
DN Tau  &  4:35:27.37 & 24:14:58.9 & 13.20 & 11.44 &   8.02 &   GN Tau B  &  4:39:20.90 & 25:45:02.1 & 17.34 & 13.76 &  8.06\\
IRAS04325+2402 &  4:35:35.39 & 24:08:19.4 &    &    & 11.60 &   J04393364p2359212& 4:39:33.64 & 23:59:21.1 & 18.28 & 15.72 & 10.28 \\
CoKu Tau3 A &  4:35:40.93 & 24:11:08.7 & 18.09 & 14.58 &   8.41 &   IRAS04365p2535 &  4:39:35.19 & 25:41:44.7 &    &    & 10.84 \\
CoKu Tau3 B &  4:35:40.93 & 24:11:08.7 & 18.09 & 14.58 &   8.41 &   ITG 15  &  4:39:44.88 & 26:01:52.7 & 20.34 & 16.62 &  8.95 \\
KPNO-8  &  4:35:41.84 & 22:34:11.6 & 20.41 & 18.04 &  11.99 &   CFHT-4  &  4:39:47.48 & 26:01:40.7 &    & 19.09 & 10.33 \\
XEST08-033 &  4:35:42.03 & 22:52:22.6 & 19.37 & 16.13 &  10.00 &   IRAS04368p2557 &  4:39:53.89 & 26:03:11.0 &    &    & 12.00 \\
J04354526+2737130& 4:35:45.26 & 27:37:13.1 &    & 19.43 & 13.71 &   IC 2087 IRS &  4:39:55.74 & 25:45:02.0 &  0.00 & 19.15 &  6.28\\
HQ Tau  &  4:35:47.33 & 22:50:21.6 & 13.04 & 11.06 & 7.14 &   J04400067p2358211& 4:40:00.67 & 23:58:21.1 & 19.45 & 17.35 & 11.48 \\
KPNO-15  &  4:35:51.10 & 22:52:40.1 & 17.78 & 15.24 &  10.01 &   CFHT-17  &  4:40:01.74 & 25:56:29.2 &    &    & 10.76 \\
IRAS04370+2559 &  4:40:08.14 & 26:05:26.54&    & 19.30 & 9.10 &  DS Tau  &  4:47:48.59 & 29:25:11.2 & 11.67 & 11.00 &   8.04 \\
J04403979+2519061& 4:40:39.79 & 25:19:06.1 & 20.21 & 17.87 & 10.24 &  J04484189+1703374& 4:48:41.90 & 17:03:37.4 &    & 19.64 & 12.49 \\
JH223  &  4:40:49.50 & 25:51:19.1 & 16.62 & 14.26 &   9.49 &  UY Aur A  &  4:51:47.37 & 30:47:13.4 & 12.47 & 11.35 &   7.24 \\
Haro 6-32 &  4:41:04.24 & 25:57:56.1 & 18.47 & 15.36 & 9.95 &  UY Aur B  &  4:51:47.37 & 30:47:13.4 & 12.47 & 11.35 &   7.24 \\
IW Tau A  &  4:41:04.70 & 24:51:06.2 & 14.39 & 11.94 &   8.28 &  IRAS04489p3042 &  4:52:06.68 & 30:47:17.5 &    &    & 10.38 \\
IW Tau B  &  4:41:04.70 & 24:51:06.2 & 14.39 & 11.94 &   8.28 &  St34  &  4:54:23.68 & 17:09:53.4 & 15.56 & 14.15 &   9.79 \\
ITG 33 A  &  4:41:08.26 & 25:56:07.4 & 20.43 & 18.63 & 11.09 &  GM Aur  &  4:55:10.98 & 30:21:59.5 & 13.84 & 11.12 &   8.28 \\
ITG 34  &  4:41:10.78 & 25:55:11.6 &    & 19.37 & 11.45 &  J04552333+3027366& 4:55:23.33 & 30:27:36.6 &  0.00 & 18.63 &  11.97 \\
IRAS04381p2540 &  4:41:12.67 & 25:46:35.4 &    &    & 11.54 &  LkCa 19  &  4:55:36.95 & 30:17:55.3 & 11.67 & 10.49 &   8.15 \\
CoKu Tau/4 &  4:41:16.81 & 28:40:00.0 & 13.99 & 12.10 &   8.66 &  J04554046+3039057& 4:55:40.46 & 30:39:05.7 & 19.07 & 17.08 &  11.77 \\
ITG 40  &  4:41:24.64 & 25:43:53.0 &    &    &  11.75 &  J04554535+3019389& 4:55:45.35 & 30:19:38.9 & 17.64 & 15.49 & 10.46 \\
IRAS04385+2550 &  4:41:38.82 & 25:56:26.5 & 18.84 & 15.54 &  9.65 &   AB Aur  &  4:55:45.82 & 30:33:04.3 &  7.13 &  7.03 &   4.23 \\
J04414489+2301513& 4:41:44.90 & 23:01:51.4 &    & 20.55 & 13.16 &  J04554757+3028077& 4:55:47.57 & 30:28:07.7 & 17.26 & 15.17 &   9.98 \\
J04414565+2301580& 4:41:45.65 & 23:01:58.0 & 16.73 & 14.24 &  9.85 &  J04554801+3028050& 4:55:48.01 & 30:28:05.5 & 20.47 & 16.66 &  12.15 \\
J04414825+2534304& 4:41:48.25 & 25:34:30.4 & 20.33 & 19.29 &  12.22 &  XEST26-052  & 4:55:48.20 & 30:30:16.0 & 17.67 & 15.80 &  10.95 \\
LkH$\alpha$ 332/G2 A &  4:42:05.48 & 25:22:56.2 & 15.55 & 11.99 &   8.23 &  J04554969+3019400& 4:55:49.69 & 30:19:40.0 & 20.14 & 17.84 &  11.86 \\
LkH$\alpha$ 332/G2 B &  4:42:05.48 & 25:22:56.2 & 15.55 & 11.99 &   8.23 &  J04555288+3006523& 4:55:52.89 & 30:06:52.3 & 17.87 & 15.80 & 10.73 \\
LkH$\alpha$ 332/G1 A &  4:42:07.32 & 25:23:03.2 & 13.80 &  9.84 &   7.95 &  XEST26-062  & 4:55:56.05 & 30:36:20.9 & 15.79 & 12.94 &  9.27 \\
LkH$\alpha$ 332/G1 B &  4:42:07.32 & 25:23:03.2 & 13.80 &  9.84 &   7.95 &  J04555636+3049374& 4:55:56.37 & 30:49:37.5 & 18.32 & 16.17 & 11.09 \\
V955 Tau A &  4:42:07.77 & 25:23:11.8 & 14.90 &  0.00 &   9.43 &  SU Aur  &  4:55:59.38 & 30:34:01.5 &  9.88 &  8.83 &   5.99 \\
V955 Tau B &  4:42:07.77 & 25:23:11.8 & 14.90 &  0.00 &   9.43 &  HBC 427  &  4:56:02.01 & 30:21:03.7 & 12.08 & 10.67 &   8.13 \\
CIDA-7  &  4:42:21.01 & 25:20:34.3 & 18.84 & 16.22 &  10.17 &  J04574903+3015195& 4:57:49.03 & 30:15:19.5 &    &    & 11.48 \\
DP Tau  &  4:42:37.69 & 25:15:37.4 & 16.71 & 14.03 &   8.76 &  V836 Tau  &  5:03:06.59 & 25:23:19.7 & 14.25 & 12.13 &  8.60 \\
GO Tau  &  4:43:03.09 & 25:20:18.7 & 16.81 & 14.24 &   9.33 &  CIDA-8  &  5:04:41.39 & 25:09:54.4 & 17.11 & 14.64 &  9.60 \\
CIDA-14  &  4:43:20.23 & 29:40:06.0 & 17.01 & 14.96 &   9.41 &  CIDA-9
&  5:05:22.86 & 25:31:31.2 & 17.04 & 14.09 &  11.16 \\
IRAS04414+2506 &  4:44:27.13 & 25:12:16.4 & 19.27 & 17.00 &  10.76&  CIDA-10  &  5:06:16.74 & 24:46:10.2 & 16.38 & 14.25 &  9.82 \\
IRAS04428p2403 &  4:45:54.82 & 24:08:43.5 & 17.67 & 15.66 &  13.19 &  CIDA-11  &  5:06:23.32 & 24:32:19.9 & 15.49 & 13.82 &  9.46 \\
RXJ04467+2459 &  4:46:42.59 & 24:59:03.1 & 18.29 & 15.85 &  15.58 &  RXJ05072+2437 &  5:07:12.07 & 24:37:16.4 & 14.32 & 12.07 & 9.30 \\
DQ Tau  &  4:46:53.05 & 17:00:00.1 & 16.28 & 12.45 &   7.98 &  RW Aur A  &  5:07:49.53 & 30:24:05.0 & 10.62 & 10.06 &  7.02 \\
Haro 6-37 A &  4:46:58.97 & 17:02:38.1 & 14.93 & 12.69 &   7.31 &  RW Aur B  &  5:07:49.53 & 30:24:05.0 & 10.62 & 10.06 &  7.02 \\
Haro 6-37 B &  4:46:59.09 & 17:02:40.3 & 14.93 & 12.69 &   7.31 &  CIDA-12  &  5:07:56.56 & 25:00:19.6 & 15.63 & 14.26 &  10.40 \\
DR Tau  &  4:47:06.20 & 16:58:42.8 & 12.08 & 10.68 &   6.87 \\
\end{longtable}
}
\end{center}

\end{landscape}

To the south, most pre-main sequence stars are in and around
L1551. In addition to L1551 IRS5 (see below), the protostar
L1551 NE and a few deeply embedded T Tauri stars (HL Tau, XZ Tau,
and HH30 IRS) form a close group of pre-main sequence stars.
A few stars lie in the larger but less dense L1556 cloud to the
east, with only a few stars outside the main cloud boundaries
(Figure \ref{fig:co2}).

Despite the filamentary structure of the dark clouds, the vast
majority of young stars in Taurus-Auriga lie in several small
groups \citep[Table 4;][]{gomez93,sim97,luh06a}. Counting each
group of stars within
20\arcsec as one system, the median separation of young stellar
systems is $\sim$ 0.3 pc, roughly a factor of 3 larger than the
radius of a dense cloud core \citep{ben89}. The typical group
has 20--30 systems within a surface area of $\sim$ 1--3 pc$^2$
\citep[see also][]{gomez93}. Current radial velocity and proper
motion data are insufficient to infer whether any of these groups
are bound.

\begin{table}[p]
\vspace{-5mm}
\caption{Groups of pre-main sequence stars in Taurus-Auriga}
\begin{center}
{\small
\begin{tabular}[t]{ l l l c c c c}
\tableline
\noalign{\smallskip}
ID & $\alpha$(2000) & $\delta$(2000) & Dark Cloud & Density (pc$^{-2}$) & Radius (pc) & Number \\
\noalign{\smallskip}
\tableline
\noalign{\smallskip}
I~~~   & 4:14:13 & +28:10:50 & B209  & 4.7 & 0.5 & 22 \\
II~~~  & 4:18:39 & +28:23:55 & L1495 & 7.2 & 0.5 & 34 \\
III~~~ & 4:40:17 & +25:45:45 & HCL2  & 2.0 & 0.9 & 30 \\
IV~~~  & 4:32:29 & +24:23:00 & L1529 & 1.3 & 1.1 & 25 \\
V~~~   & 4:34:40 & +22:49:30 & L1536 & 1.8 & 1.0 & 33 \\
VI~~~  & 4:31:47 & +18:09:40 & L1551 & 1.2 & 1.1 & 25 \\
\noalign{\smallskip}
\tableline
\end{tabular}
}

\caption{Close binaries in Taurus-Auriga$^1$}
{\small
\begin{tabular}[t]{ l@{\hskip6pt} c@{\hskip6pt} c@{\hskip6pt} | l@{\hskip6pt} c@{\hskip6pt} c}
\tableline
\noalign{\smallskip}
Binary & Sep ($^{\prime \prime}$) & $\delta$K (mag) & Binary & Sep ($^{\prime \prime}$) & $\delta$K (mag) \\
\noalign{\smallskip}
\tableline
\noalign{\smallskip}
HBC 351       & 0.6 & 1.6 &         CFHT-7       & 0.2 & 0.4 \\
HBC 356/357   & 2.0 & &             V928 Tau      & 0.2 & 0.6 \\
HBC 358       & 1.6 & 0.6 &         GG Tau A      & 0.3 & 0.5--1.2 \\
V773 Tau AB   & sb  & &             GG Tau B      & 1.5 & 1.7--1.8 \\
V773 Tau AB-C & 0.1 & 1--2 &    UZ Tau W      & 0.4 & 0.5--1.1 \\
LkCa 3       & 0.5 & 0.0 &          UZ Tau E      & sb  & \\
FO Tau       & 0.2 & 0.5 &          GH Tau       & 0.3 & 0.6 \\
V410 Tau AB   & 0.1 & 2.0 &         V807 Tau      & 0.3 & 1.1 \\
V410 Tau A-C  & 0.3 & 3.0 &     GK Tau       & 2.4 & \\
DD Tau       & 0.6 & 0.7--0.9 &     IS Tau       & 0.2 & 2.6 \\
CZ Tau       & 0.3 & 0.8 &          HN Tau       & 3.1 & 3.5 \\
FQ Tau       & 0.8 & 0.1 &          IT Tau       & 2.5 & 1.6 \\
LkCa 7       & 1.1 & 0.6 &          FF Tau       & 0.1 & 1.0 \\
T Tau NS      & 0.7 & 2--6 &        HBC 412       & 0.7 & 0.0 \\
T Tau S       & 0.1 & 0.5--3.0 &    CoKu Tau/3    & 2.0 & 1.9 \\
FS Tau       & 0.3 & 2.3 &          HP Tau       & 0.1 & 2.3 \\
J4872       & 3.0 & 0.8 &           HP Tau/G3     & 0.1 & 1.4 \\
FV Tau       & 0.7 & 0.2 &          Haro 6-28     & 0.7 & 0.5 \\
FV Tau/c      & 0.7 & 1.9 &         HV Tau Aa     & 0.1 & 0.6 \\
IRAS04239+2436   & 0.3 & 0.5--1.0 &    HV Tau AB     & 4.0 & 3.8 \\
DF Tau       & 0.1 & 0.4--0.9 & IRAS04361+2547$^3$   & 0.3 & 0.4 \\
IRAS04248+2612$^2$& 4.6 & 4.6 &        VY Tau       & 0.7 & 1.5 \\
J04284263+2714039 & 0.6 & 0.9 &     GN Tau       & 0.1 & 0.5 \\
GV Tau       & 1.2 & 2.2 &          CFHT-17       & 0.6 & 1.5 \\
IRAS04263+2426   & 1.3 & 1.1 &         J04403979+2519061 & 0.1 & 1.1 \\
FW Tau       & 0.2 & 0.0 &          IW Tau       & 0.3 & 0.0 \\
DH Tau A      & 2.3 & 6.8 &         CoKu Tau/4    & 0.1 & 0.2 \\
DI Tau       & 0.1 & 2.3 &          IRAS04385+2550   & 18.9 & 3.2 \\
UX Tau A-C    & 2.7 & 2.9 &         LkH$\alpha$ 332/G1 & 0.2 & 0.6 \\
FX Tau       & 0.9 & 0.4 &          LkH$\alpha$ 332/G2 & 0.3 & 0.6 \\
DK Tau       & 2.5 & 1.3 &          V955 Tau      & 0.3 & 1.6 \\
ZZ Tau       & 0.1 & 0.9 &          Haro 6-37 Aa  & 0.3 & 2.5 \\
V927 Tau      & 0.3 & 0.3 &         Haro 6-37 AB  & 2.7 & 0.9 \\
XZ Tau       & 0.3 & 0.7 &          UY Aur       & 0.9 & 1.4 \\
HK Tau       & 2.4 & 3.4 &          RW Aur A-BC   & 1.4 & 2.3 \\
V826 Tau      & sb & &              RW Aur B-C    & 0.1 & 4.0 \\
\noalign{\smallskip}
\tableline
\noalign{\smallskip}
\multicolumn{6}{l}{\parbox{0.95\textwidth}{\footnotesize
    $^1$Spectroscopic (sb) and close binaries not resolved as distinct
    objects in 2MASS. Results are quoted to the nearest 0.1 arcsec
    (separations) and the nearest 0.1 mag ($\Delta K$). }}\\[2ex]
\multicolumn{6}{l}{\parbox{0.95\textwidth}{\footnotesize
    $^2$2MASS data show a K = 12.8 companion with a separation of
    2.6\arcsec. }}\\[1ex]
\multicolumn{6}{l}{\parbox{0.95\textwidth}{\footnotesize
    $^3$Not resolved at L by \citet{gram07}.}}\\
\end{tabular}
}
\end{center}
\end{table}

Many pre-main sequence stars are in binary or multiple systems
\citep[Table 5;][]{lei89,haas90,sim92,sim93,sim95,sim00,ghe93,
math1994,rich94,kor97, koh98,white99,mort00,kon01,woi01,rei02,
tam02,duc03b,duc04,duc06,kraus06,gram07,kono2007,ire08}.
The binary frequency among this sample is comparable to or
slightly larger than the field binary frequency. Although
most binaries are pairs of weak emission T Tauri stars or
classical T Tauri stars with bright emission lines, there are
a few mixed pairs of weak and classical T Tauri stars.  In most
binaries and multiple systems, the stars are coeval, but there
are some large age differences
\citep[e.g.,][]{har94,pra97,whi01,har03,pra03,hill04}.
Further study is needed to see whether these differences are
real, due to errors in the pre-main sequence evolutionary tracks,
or the result of different accretion histories.

\begin{figure}[!tb]
\epsscale{0.8}
\plotone{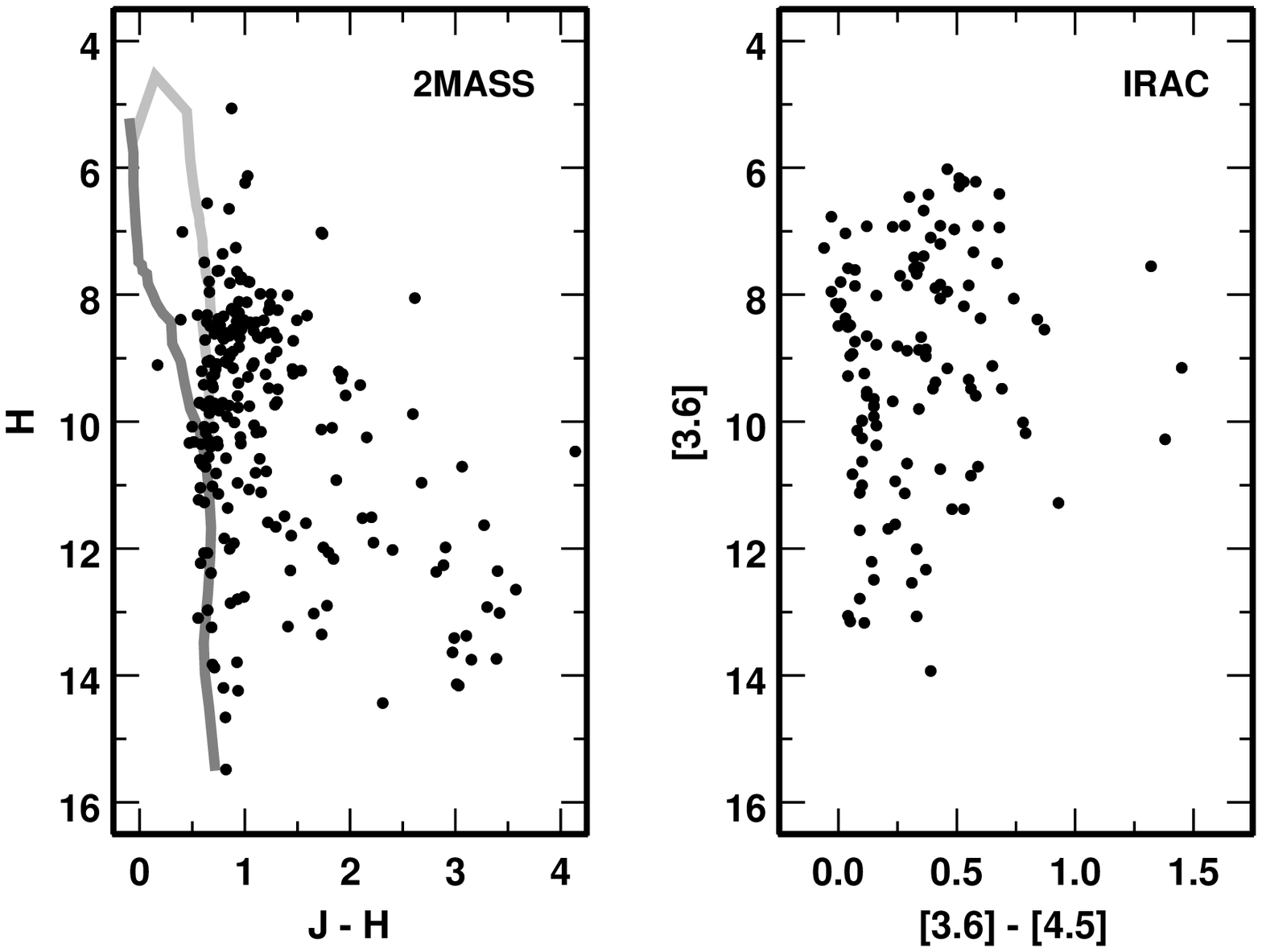}
\vspace{-12mm}
\caption{Infrared color-magnitude diagram for Taurus-Auriga pre-main
sequence stars. {\it Left panel:} Data from 2MASS.  TTS lie above
and to the right of the ZAMS (dark gray line) and the 1 Myr
isochrone (light gray line) from Siess et al. (2000).
{\it Right panel:} Data from IRAC. Most TTS have large color
excesses relative to the few TTS that lie close to the 1 Myr
isochrone.}
\label{fig:cmd}
\vspace{-1.5ex}
\end{figure}

The large sample of pre-main sequence stars in Taurus-Auriga yields
good tests of stellar evolution models. \citet{kho51} and \citet{her52}
first showed that T Tauri stars lie 1--3 mag above the main sequence
(Figure \ref{fig:cmd}). \citet[][1966]{hay65} used stellar structure
calculations to show that these stars are contracting to the main
sequence along `Hayashi tracks' with roughly constant effective
temperature \citep[see also][]{coh79}.
Later elaborations of these models provide a better understanding
of how pre-main sequence stars form and evolve \citep[e.g.][]{bar95,
bar02,sie96,sie97,sie99,tou99,white99,hart02,pal02,mon05,ber07}.

Because dynamical masses for T Tauri stars have been rare,
stellar evolution models are the best way to estimate masses
and to measure the initial mass function (IMF) for young stars.
In Taurus-Auriga, the large pre-main sequence population provides
the standard IMF for a nearby young association with stellar ages
of 1--10 Myr \citep{ken95,ito96,luh04}.
The recent discovery of brown dwarfs extends the IMF to substellar
masses \citep{bri93,bri02,luh00,luh04,luh06a,mar01,gui06,luh06b}.

From X-ray to radio wavelengths, T Tauri stars have impressive displays
of periodic and random brightness variations. In the UV, optical, and
near-IR, modulations due to hot and cold spots produce repeatable
variations with periods of a few days to almost two weeks \citep{bou88,
bou92,bou93,bou89,aud07,ste07}.  Random changes in brightness occur
on timescales of days to years \citep{ryd84a,her94,johns95,gro07a}.
In X-rays and radio emission, occasional large flares erupt with
frequencies similar to those observed in the Sun, but with luminosities
several orders of magnitude larger
\citep{oneal90,phi93,phi96,fei94,chi96,cark96,cark97,cark98,gia06,aud07}.

In addition to their variability and strong emission lines,
many T Tauri stars have considerable UV continuum
emission compared to a main sequence stellar photosphere with
the same spectral type \citep{her58,her60,her77,var60,sma64,rss76}.
Because this `veiling' is an extra source of continuum emission,
it fills in stellar absorption lines. Together with analyses of
IR excesses, measurements of stellar veiling have led to
detailed accretion disk models and fairly robust estimates of
accretion rates from the disk onto the star
\citep{ryd84b,ruc85,ken87,ken90a,ber88,bas89,ber89,har90,hart91,
gul98,john00,john01,har03,edw06}.

Robust measurements of continuum veiling and rotational modulation
of T Tauri stars prompted magnetospheric accretion models, where the
stellar magnetic field truncates the disk at 3--5 stellar radii and
channels the accretion flow onto magnetic hotspots, which rotate with
the stellar photosphere \citep{kon91,cal92,edw93,edw94,cla95,ken96,
mah98,muz98,muz03,gul00, bou99,oli00,ber01,bou03,sym05,eis05,osu05}.
Magnetic field measurements provide some support for this picture
\citep{bas92,john99,johns04}.  This picture also provides physical
mechanisms to power stellar jets and Herbig Haro flows
\citep{shu94,kud97,hir97,sok03,kras03},
and to understand the distribution and evolution of rotational
periods \citep{col93,yi94,yi95,all96,kri97,sta03,bro06,gro07b}.

\section{Herbig-Haro Flows and Molecular Outflows}

In the 1950's, \citet{herb51} and \citet[][1953]{haro52} discovered
bright knots of nebulosity in several star-forming regions.  Besides
intense H~I emission lines, the knots showed strong emission from [S~II]
$\lambda\lambda$4068, 4076, 6717, 6731; [O~I] $\lambda\lambda$6300, 6363;
and [O~II] $\lambda\lambda$3726, 3729. Sensitive spectra also revealed
weak emission from Ca II and [Fe II].  Although early observations made
a clear association of these `Herbig-Haro' (HH) objects with young stars
in Taurus-Auriga and other molecular clouds \citep{haro53,herbig74,schwa75,
elias78}, modern data demonstrate that HH objects, jets, and outflows are
an important -- perhaps necessary -- part of low mass star formation
\citep[e.g.,][]{mufr83,mundt84,stro86,bal07}.

Table 6 lists the currently known jets and HH flows in Taurus-Auriga.
Most jets have been discovered using the optical [O~I], H$\alpha$,
[N~II] $\lambda\lambda$ 6548, 6583, and [S~II] $\lambda\lambda$ 6717,
6731 emission lines (e.g., G\'omez et al. 1997; Devine et al 1999a,
1999b).  In regions of large extinction, near-IR transitions of H$_2$
and [Fe~II] are also important \citep{dav02,dav03}.  Sensitive imaging
surveys covering most of the clouds have detected flows associated
with the most deeply embedded young stars (e.g., L1551 IRS5, L1527
IRS, and  IRAS04248+2612) and optically visible T
Tauri stars with clear evidence for a circumstellar disk (e.g., T Tau,
DG Tau, and RW Aur).  Despite
sensitive searches, there are no weak emission T Tauri stars with an
HH outflow or a jet. Thus, these surveys suggest a clear link between
accretion and outflows.

\begin{figure}[!ht]
\centering
\plotone{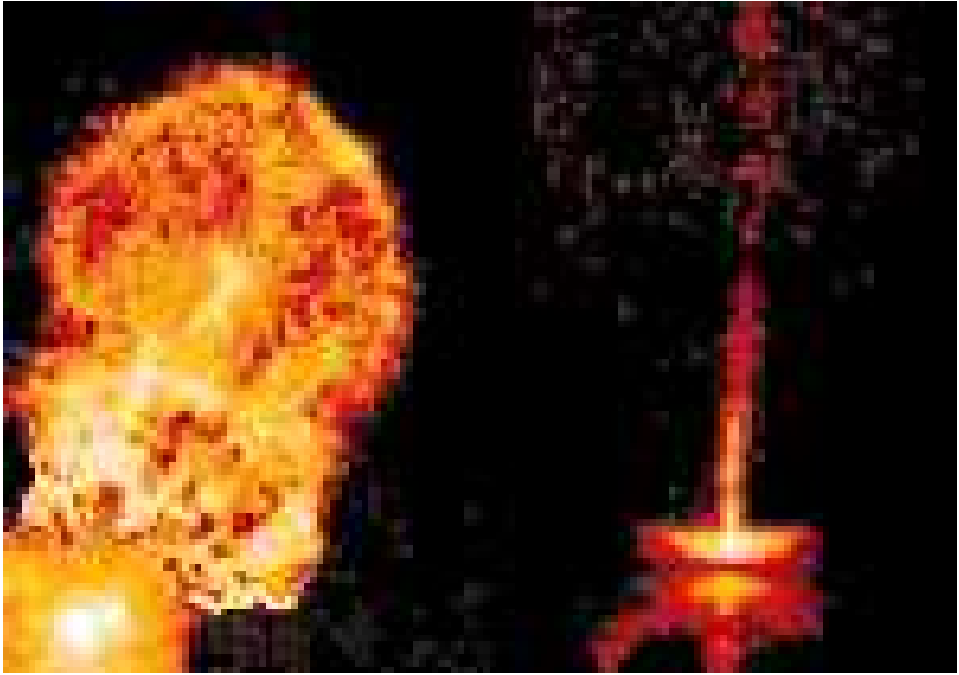}
\caption{HST WFPC2/F675W images of XZ Tau (left; Krist et al. 1999)
and HH 30 (right; Burrows et al. 1996).  The F675W filter includes 
the [S II] $\lambda$$\lambda$6717,6731, H$\alpha$ $\lambda$6563,
and [O I] $\lambda$6300 lines. The XZ Tau bubble extends for
800--900 AU from the central binary. The HH 30 jet has a length
of 1300--1400 AU from the disk midplane.
\label{fig:XZHH}
}
\end{figure}

Figure \ref{fig:XZHH} illustrates some of the amazing variety of jets
among Taurus-Auriga young stars. In HH 30 and other jet sources, the
jets typically have opening angles of roughly 5 degrees and linear
dimensions ranging from a few tens of AU to about 0.2 pc (e.g., Kepner
et al. 1993; Ray et al. 1996; Burrows et al. 1996; Krist et al. 1997,
1999; Lavalley et al. 1997; Fridlund \& Liseau 1998; Stapelfeldt et al.
1999, 2003; Dougados et al.  2000; Woitas et al. 2002a,b, 2005; Bacciotti
et al. 2002; Pyo et al. 2003, 2005; Coffey et al. 2004a, b; Hartigan
et al. 2004; Fridlund et al. 2005; McGroarty et al. 2007). When resolved,
the lateral dimensions
are $\sim$ 10--20 AU.  Jets come in two broad classes, nebulous knots of
HH objects arranged like `beads on a string' and continuous narrow jets.
Few jets are perfectly straight.  Many jets curve slightly, while others
have a wavy or sinusoidal shape.  This morphology may indicate that the
source of the jet wobbles, precesses, or is part of a binary system.
Both classes have a bipolar (or sometimes unipolar) morphology, with the
jet or HH object lying roughly along the apparent rotation axis of the
circumstellar disk surrounding the T Tauri star.

Some jets and HH objects have impressive bow shocks, arc-shaped nebulae
formed as ejected material plows through the molecular cloud. In XZ Tau,
{\it HST} data show a clear expansion of the bubble-like structure
surrounding the HH knots (Krist et al. 1999). Often, the HH objects
also show a clear proper motion away from the central young star (e.g.,
Cudworth \& Herbig 1979). The
expansion velocities derived from the proper motions, $\sim$ a few
hundred km s$^{-1}$, typically agree with expansion velocities derived
from the profiles of the [O~I] and [S~II] emission lines
(e.g., G\'omez de Castro 1993; Hirth et al. 1994a, 1994b, 1997;
B\"ohm \& Solf 1994; Bacciotti et al. 1996; Eisl\"offel \& Mundt 1998;
Solf \& B\"ohm 1999; Takami et al. 2002; Sun et al. 2003).

Many HH flows are associated with reflection nebulae. In L1551 IRS5
and many other jets, impressive reflection nebulae surround the jet,
suggesting that the jet flows through a cavity in the surrounding
molecular gas. Other TTS have bowl-shaped or C-shaped reflection
nebulae. In HH 30 (Fig. 8), the beautiful bipolar jet is associated
with a bipolar bowl-shaped nebula almost perfectly bisected by a
dark band of obscuring material. This image closely resembles the
predicted images of illuminated disks observed edge-on (Whitney \&
Hartmann 1992, 1993); the upper half of the nebula is the near side
of the disk, while the lower half is the far side of the disk.

In addition to large-scale jets, discoveries of `microjets' in DG Tau
(Kepner et al. 1993) and RW Aur \citep{baccio96} provide important
details on the structure and evolution of outflows from young stars.
In DG Tau, images with high spatial resolution show several knots
lying inside an expanding bow shock, with structure similar to XZ Tau
but on scales of 1--2 arcsec instead of 5--10 arcsec
\citep{lavalley97,laval00,baccio00,dougados00,tak02,tak04b,pyo03,
gud05,gud08}.
Small wiggles in the jet and the spacing and apparent trajectory of
the knots suggest a precessing outflow source with recurrent ejections
on timescales $\sim$ 8 yr.
Sophisticated models show that a precessing jet accounts for the
excitation, kinematics, and morphology of the emission lines
\citep{raga01,cerq04,mass05}.

On the largest scales, continued improvement of large-format optical
CCD and infrared HgCdTe and InSb cameras enable amazing detections of
beautiful new jets and many faint and distant components of known jets
and HH objects (e.g., Eiroa et al. 1994; Alten et al. 1997;
G\'omez et al. 1997; Lucas \& Roche 1998; Eisl\"offel \& Mundt 1998;
Aspin \& Reipurth 2000; Magakian et al. 2002; McGroarty et al 2007).
Imaging of giant HH
flows often reveals quasi-periodically spaced ejections with impressive
point symmetry (e.g., Reipurth et al. 1997, 2000; Devine et al 1999a,
1999b; Sun et al. 2003; McGroarty \& Ray 2004; Wang et al. 2004). In
IRAS 04248+2612, the sinusoidal outflow trajectory resembles the
structure in DG Tau, suggesting that precessing jets may explain
the basic geometry of most jets (G\'omez et al. 1997). Because this
source is a known binary (Padgett et al. 1999), binary motion might
also produce a sinusoidal outflow.

The young stars in Taurus-Auriga also drive impressive large-scale
mm and cm outflows.  After \citet{snell80} discovered the bipolar
CO outflow in L1551 IRS5, high velocity molecular gas was observed
around many T Tauri stars \citep[e.g.,]{edsne82,kut82,bala83,heyer87,
moriarty87,moriarty92,moriarty95,moriarty91,myers88,morisnell88,terebey89,
fukui89}. At about the same time, \citet{cohen82} detected large-scale
ionized outflows associated with 4 young stars in Taurus-Auriga. Larger
surveys with the VLA continued to reveal ionized outflows associated with
other T Tauri stars, while more sensitive, high resolution observations
revealed the fine details of the outflows \citep[e.g.,][]{rodr83,bieg84,
sch86,rodr94}.

Modern radio observations concentrate on resolving small-scale
structures within the outflow, deriving the orientation of the
outflow relative to the disk and optical jet, and measuring the
physical parameters and chemistry of the outflow. In Taurus-Auriga,
the projected lengths, $\sim$ 0.1--1 pc, and velocities, $\sim$
10--100 km s$^{-1}$, roughly span the
range observed in HH objects
and collimated jets \citep{bac99,ric00,arcegood01}. Outflow rates,
$\sim 10^{-6}$ $M_{\odot}$ yr$^{-1}$ or less, are also similar to
those derived from optical and near-IR data and comfortably less
than the typical inflow rates within the molecular cloud
\citep{bontemps96}. The radio data show that the molecular outflows
are much cooler, $\sim$ 20--100 K, than optical jets, as expected for
molecular gas. These regions have a rich chemistry, with an impressive
array of charged and neutral molecules \citep[e.g.,][]{hog98,spin00}

While most protostars in Taurus-Auriga drive large-scale molecular outflows
\citep[Table 6;][]{heyer87,terebey89,moriarty91}, only one protostar --
04166+2706 -- drives a well-collimated, high velocity, bipolar molecular
outflow \citep[`molecular jet';][]{taf04}. This system is similar to other
molecular jets in Ophiuchus and Perseus, with an outflow velocity of
at least 50 km s$^{-1}$ and a high degree of symmetry between the
blue-shifted and red-shifted material. Because this system has weak
optical [S~II] emission \citep{gomez97}, deeper optical and near-IR
images might reveal a well-collimated optical jet associated with the
molecular gas.

\begin{landscape}
\begin{center}
{ \scriptsize
\begin{longtable}{ l @{\hskip6pt} l@{\hskip6pt} l@{\hskip6pt} l@{\hskip6pt} l@{\hskip6pt} l@{\hskip-12pt} r@{\hskip6pt} l}

\caption{Herbig Haro Flows in Taurus-Auriga\label{table4a}}\\
\noalign{\smallskip}
\tableline
\noalign{\smallskip}
HH & Other ID & $\alpha$(2000) & $\delta$(2000) & Driving Source &
Description & CO Outflow? & References \\
\noalign{\smallskip}
\tableline
\noalign{\smallskip}
\endfirsthead

\caption{Herbig Haro Flows in Taurus-Auriga (continued)}\\
\noalign{\smallskip}
\tableline
\noalign{\smallskip}
HH & Other ID & $\alpha$(2000) & $\delta$(2000) & Driving Source &
Description & CO Outflow? & References \\
\noalign{\smallskip}
\tableline
\noalign{\smallskip}
\endhead

\noalign{\smallskip}
\tableline
\endfoot

\noalign{\smallskip}
\tableline
\noalign{\smallskip}
\multicolumn{8}{l}{\parbox{21cm}{
\footnotesize
References:
1- \cite{alten97}; 2- \cite{gomez97};
3- \cite{myers88}; 4- \cite{terebey89};
5- \cite{moriarty92}; 6- \cite{sun03};
7- \cite{mcgroray04}; 8- \cite{gomezdec93};
9- \cite{hirth94a}; 10- \cite{dougados00};
11- \cite{har04}; 12- \cite{movmaga90};
13- \cite{eislmun98}; 14- \cite{bontemps96};
15- \cite{rei97}; 16- \cite{sobo99};
17- \cite{schwa75}; 18- \cite{buhr86};
19- \cite{boso94}; 20- \cite{asprei00};
21- \cite{knapp77}; 22- \cite{edsne82};
23- \cite{calvet83}; 24- \cite{lada85};
25- \cite{levreault88}; 26- \cite{cabber92};
27- \cite{moriarty87}; 28- \cite{lari99};
29- \cite{bur94}; 30- \cite{sto08};
31- \cite{mundt84};
32- \cite{stro86}; 33- \cite{woi02a};
34 - \cite{krist98}; 35- \cite{mundt91};
36- \cite{rei00}; 37- \cite{arcegood01};
38- \cite{dav03}; 39- \cite{mufr83};
40- \cite{mundt87}; 41- \cite{lavalley97};
42- \cite{kep93}; 43- \cite{baccio00};
44- \cite{baccio02}; 45- \cite{herbig74};
46- \cite{dev99a}; 47- \cite{elias78};
48- \cite{hirth97}; 49- \cite{edsne84};
50- \cite{fukui89}; 51- \cite{grahe90};
52- \cite{garnav92}; 53- \cite{dev99b};
54- \cite{dav95}; 55- \cite{sto88};
56- \cite{rod89}; 57- \cite{frilis94};
58- \cite{fridli98}; 59- \cite{necsta87};
60- \cite{camp88}; 61- \cite{yata92};
62- \cite{liseau05}; 63- \cite{fridl05};
64- \cite{snell80}; 65- \cite{bala83};
66- \cite{morisnell88}; 67- \cite{friwhi89};
68- \cite{moriarty91}; 69- \cite{rod95};
70- \cite{momose98}; 71- \cite{frikne93};
72- \cite{fridl93}; 73- \cite{cudhe79};
74- \cite{moriarty95}; 75- \cite{mundt88};
76- \cite{mundt90}; 77- \cite{lopez95};
78- \cite{burrows96}; 79- \cite{ray96};
80- \cite{sta96}; 81- \cite{brugel81};
82- \cite{cohen81}; 83- \cite{raga97};
84- \cite{lopez96}; 85- \cite{sta95};
86- \cite{bacceis99}; 87- \cite{torre87};
88- \cite{monin96}; 89- \cite{cohejo87};
90- \cite{krist97}; 91- \cite{haro53};
92- \cite{maga02}; 93- \cite{wang01};
94- \cite{heyer87}; 95- \cite{tamura96};
96- \cite{woilei98}; 97- \cite{sta03}
98- \cite{eiroa94}; 99- \cite{chan96};
100- \cite{sta99}; 101- \cite{mueisl98};
102- \cite{hirth94b}; 103- \cite{baccio96};
104- \cite{woi02b}; 105- \cite{lopez03};
106- \cite{coffey04}; 107- \cite{woi05}
}}\\
\endlastfoot

  362&                    &   4:04:23.0 &    26:20:41 &   04016+2610?            & small knots                  & yes & 1,2,3,4,5 \\
  361&                    &   4:04:34.9 &    26:21:44 &   04016+2610?            & small knots                  &     & 1,2 \\
  360&                    &   4:04:43.0 &    26:19:00 &   04016+2610?            & small knots                  &     & 1,2 \\
  701&                    &   4:12:16.4 &    28:50:15 &                          & knots                        &     & 6 \\
  829&                    &   4:14:03.0 &    28:25:36 &   CW Tau                 & knot                         &     & 7  \\
  828&                    &   4:14:10.3 &    28:14:54 &   CW Tau                 & knot                         &     & 7  \\
  827&                    &   4:14:15.1 &    28:03:55 &   CW Tau                 & knot                         &     & 7  \\
  220&  CW Tau  jet       &   4:14:16.9 &    28:10:59 &   CW Tau                 & bipolar jet                  &     & 7,8,9,10,11 \\
  826&                    &   4:14:17.8 &    28:10:40 &   CW Tau                 & knot                         &     & 7  \\
     &   DD Tau A jet     &             &             &   DD Tau A               & [O II] jet                   &     & 11 \\
  156&  CoKu Tau-1 jet    &   4:18 51.5 &    28:20:28 &   CoKu Tau-1             & bipolar jet                  &     & 12,13 \\
  390&                    &   4:19 40.8 &    27:15:53 &   04166+2706?            & small knots                  & yes & 2,14 \\
  391&                    &   4:19 56.3 &    27:09:26 &   04169+2702?            & small knots/HH jet?          & yes & 2,5,14 \\
  392&                    &   4:20 54.3 &    26:59:52 &   04181+2655/54?         & small knots                  & yes & 2,5,14 \\
  355&                    &   4:21:43.6 &    19:50:42 &   T Tauri S              & bipolar HH jet               &     & 15,16 \\
  155&  HH 1555           &   4:21:57.1 &    19:32:07 &   T Tauri                & bipolar jet                  & yes & 5,13,16,17,18,19,20,21,22,23,24,25,26,27,28 \\
  998&                    &   4:21:57.8 &    28:26:35.9 & RY Tauri
   & small knots/bipolar jet      &     & 30 \\
  255&  Burnham's Nebula  &   4:21:59.4 &    19:31:56 &   T Tauri S              & HH jet                       &     & 16,18,19,29 \\
  157&  Haro 6-5B jet     &   4:22:00.9 &    26:57:38 &   Haro 6-5B(FS Tau B)    & HH jet                       &     & 13,31,32,33,34,35 \\
  276&                    &   4:22:07.3 &    26:57:26 &                          & HH jet                       &     & 13,32 \\
  300&  HH 300 [FeII] jet &   4:25:23.0 &    24:23:20 &   04239+2436             & HH jet/bow-shock shape knots & yes & 2,5,15,36,37,38 \\
  702&                    &   4:26:35.6 &    25:57:55 &                          & knots                        &     & 6 \\
     & DF Tau jet         &             &             &   DF Tau                 & [O II] jet                   &     & 11 \\
  838&                    &   4:26:56.4 &    26:05:58 &   DG Tau B?              & knot                         &     & 7 \\
  159&  DG Tau B jet      &   4:27:02.0 &    26:05:42 &   DG Tau B               & jet                          &     & 7,13,39,40,41 \\
  836&                    &   4:27:13.5 &    26:04:16 &   DG Tau B               & knot                         &     & 7 \\
  839&                    &   4:27:43.8 &    26:04:35 &   DG Tau B?              & knot                         &     & 7 \\
  837&                    &   4:27:44.8 &    26:00:49 &   DG Tau B               & knot                         &     & 7 \\
  158&  DG Tau jet        &   4:27:04.6 &    26:06:16 &   DG Tau                 & bipolar jet                  &     & 7,10,13,31,38,39,40,42,43,44 \\
  830&                    &   4:27:37.3 &    26:12:27 &   DG Tau                 & knot                         &     & 7 \\
  31&                     &   4:28:18.4 &    26:17:41 &   04248+2612             & jet/knots                    & yes & 2,5,32,45 \\
  410&                    &   4:28:13.0 &    24:19:02 &   Haro 6-10 IR comp      & bright knots/bow shock       &     & 46      \\
  184& Haro 6-10/HH       &   4:29:23.6 &    24:33:01 &   Haro 6-10 IR comp      & compact knot                 &     & 2,31,46,47,48 \\
     & Haro 6-10 jet      &   4:29:24.4 &    24:33:02 &   Haro 6-10              & small bipolar jet            & yes & 4,25,46,49,50,28 \\
  412&                    &   4:29:47.9 &    24:37:10 &   Haro 6-10 IR comp      & diffuse emission             &     & 46 \\
  411&                    &   4:30:16.9 &    24:42:42 &   Haro 6-10 IR comp      & H$\alpha$ filament           &     & 46 \\
  414&                    &   4:29:30.3 &    24:39:54 &   04264+2433             & bipolar jet                  &     & 46 \\
  413&                    &   4:29:53.0 &    24:38:12 &   04264+2433             & bow-shock                    &     & 46 \\
 393&                     &   4:30:50.6 &    24:41:25 &   ZZ Tau?                & small knot                   & yes & 2,50 \\
 256& GH 1                &   4:30:53.2 &    17:59:07 &   L1551 IRS 5,HH 30 IRS? & small faint knots            &     & 51,52,53 \\
 257& GNG 17              &   4:31:00.7 &    18:00:42 &   L1551 IRS 5            & faint knot                   &     & 52 \\
 258& GH 2-8              &   4:31:04.8 &    18:03:32 &   L1551 IRS 5            & faint knots                  &     & 51 \\
 260& GNG 1               &   4:31:27.0 &    18:06:55 &   L1551 IRS 5            & faint knot                   &     & 52,54 \\
 261& SH 219/220          &   4:31:30.0 &    18:06:53 &   L1551 IRS 5            & knots                        &     & 51,52,55,56 \\
 154& L1551 IRS5 jet      &   4:31:33.8 &    18:08:02 &   L1551 IRS 5            & short jet                    & yes & 5,24,26,27,32,34,38,39,51,52,54,55,57,58,59 \\
    &                     &             &             &                          &                              &     & 60,61,62,63,64,65,66,67,68,69,70,71 \\
 264& GNG 4               &   4:31:18.4 &    18:06:16 &   L1551 IRS 5?           & knots                        &     & 51,52,54,55 \\
 262& GH 9/10             &   4:32:01.1 &    18:11:24 &   L1551 IRS 5,L1551 NE?  & faint knots                  &     & 51,52,53,56 \\
 28&                      &   4:31:07.0 &    18:03:23 &   L1551 NE               & bright bow shock             &     & 32,45,51,52,53,54,55,72,73 \\
 29&                      &   4:31:27.0 &    18:06:23 &   L1551 NE               & bright bow shock             &     & 32,45,51,52,53,54,55,72,73 \\
 454&                     &   4:31:42.7 &    18:08:19 &   L1551 NE               & bipolar HH jet               & yes & 53,74 \\
 259& SH 229 group        &   4:31:14.1 &    18:04:02 &   L1551 NE?              & knots/HH jet                 &     & 51,52,53,55,56 \\
 286&                     &   4:32:41.7 &    18:16:36 &   L1551 NE?,L1551 IRS 5? & knot                         &     & 53 \\
 265& GNG 25              &   4:31:14.9 &    18:11:59 &                          & knot                         &     & 52,53 \\
 263& SH 214, GNG 3       &   4:31:23.4 &    18:07:48 &   HH 30 IRS?             & small knots                  &     & 51,52,54,55 \\
 30&                      &   4:31:37.6 &    18:12:26 &   HH 30 IRS              & bipolar jet                  &     & 34,35,40,45,51,53,75,76,77,78,79,80,81,82,83,84 \\
 150& HL Tau jet          &   4:31:38.5 &    18:13:59 &   HL Tau                 & bipolar jet                  & yes & 22,23,24,35,75,76,77,79,84,85,86,87,88 \\
 153& HL Tau H$\alpha$ jet&   4:31:39.4 &    18:13:39 &   HL Tau                 & H$\alpha$ jet,[S II]knot     &     & 75,76 \\
 151& HL Tau VLA 1 jet    &   4:31:39.6 &    18:14:08 &   HL Tau VLA             & bipolar jet                  &     & 32,34,35,40,75,76,89 \\
 266& GNG 24              &   4:31:52.3 &    18:16:50 &   HL Tau?                & knot                         &     & 52,77,84 \\
 152& XZ Tau jet          &   4:31:40.1 &    18:13:58 &   XZ Tau                 & bipolar jet                  &     & 32,48,75,76,87,90 \\
    & Haro 6-13 H$\alpha$ jet&          &             &   Haro 6-13              & unipolar jet               &     & 32 \\
 319& Haro 6-19           &   4:32:41.1 &    24:21:46 &   FY Tau?,FZ Tau?        & knots                        &     & 6,91,92 \\
    & UZ Tau E jet        &             &             &   UZ Tau E               & [O I],[N II],[SII] jet       &     & 11,48 \\
 394&                     &   4:33:12.3 &    22:55:10 &   04302+2247?            & HH knots                     & yes & 2,5,15 \\
 467&                     &   4:33:32.9 &    24:20:27 &   GK Tau                 & knot/extended emission       &     & 20 \\
 466&                     &   4:33:35.4 &    24:21:32 &   GK Tau                 & knots                        &     & 20 \\
 468&                     &   4:33:37.6 &    24:21:43 &   GK Tau                 & diffuse knot                 &     & 20 \\
    & HN Tau jet          &             &             &   HN Tau                 & [O I],[N II],[SII] jet       &     & 11,48 \\
 434&                     &   4:34:13.2 &    23:09:29 &   04325+2402(L1535)?     & knots                        & yes & 3,5,93,94,95 \\
 435&                     &   4:34:15.1 &    23:08:08 &   04325+2402(L1535)?     & bow shock                    & yes & 5,93 \\
 436&                     &   4:34:20.3 &    23:08:40 &   04325+2402(L1535)?     & elongated knot               &     & 93 \\
 703&                     &   4:35:01.9 &    23:38:58 &                          & nebula                       &     & 6 \\
 230& DO Tau jet          &   4:38:28.6 &    26:10:50 &   DO Tau                 & bipolar knots                &     & 7,9,48 \\
 833& HV Tau C jet        &   4:38:44.0 &   26:14:42  &   HV Tau                 & [O I],[SII] jet              &     & 7,41,96,97 \\
 834&                     &   4:39:05.9 &   26:03:23  &   HV Tau?                & knot                         &     & 5          \\
 832&                     &   4:39:02.0 &    26:12:21 &   DO Tau                 & knot                         &     & 7 \\
 831&                     &   4:39:13.2 &    26:13:48 &   DO Tau                 & knot                         &     & 7 \\
 704&                     &   4:38:45.1 &    25:18:14 &                          & knots/nebulae                &     & 6 \\
 705&                     &   4:39:06.7 &    26:20:30 &                          & nebula                       &     & 6 \\
 706&                     &   4:39:11.4 &    25:27:19 &                          & nebula                       &     & 6 \\
 192&                     &   4:39:47.7 &    26:03:27 &   04368+2557?            & small knot                   & yes & 4,5,14,95,98,99 \\
 395&                     &   4:40:08.7 &    25:46:44 &   04369+2539(IC 2087)    & small knots                  & yes & 2,22,28,94 \\
 408&                     &   4:41:38.9 &    25:56:26 &   Haro 6-33              & bipolar knots                &     & 100 \\
 231& DP Tau jet          &   4:42:37.6 &    25:15:38 &   DP Tau                 & bipolar HH jet               &     & 9,13,48,101 \\
 386& UY Aur jet          &   4:51:47.3 &    30:47:15 &   UY Aur                 & bipolar microjet             &     & 47,48\\
 835&                     &   5:07:30.4 &    30:27:11 &   RW Aur                 & knot                         &     & 7 \\
 229& RW Aur jet          &   5:07:49.5 &    30:24:07 &   RW Aur                 & bipolar HH jet               &     & 7,10,38,47,48,101,102,103,104,105,106,107\\
\end{longtable}
}
\end{center}
\end{landscape}

\begin{table}
\caption{Molecular Outflows without Detected Jets/HH Objects in Taurus-Auriga}
\smallskip
\begin{center}
{\small
\begin{tabular}[t]{ l l l l l}
\tableline
\noalign{\smallskip}
ID & Other ID & $\alpha$(2000) & $\delta$(2000) & References \\
Description & CO Outflow? & References \\
\noalign{\smallskip}
\tableline
\noalign{\smallskip}
04191+1523 &        & 4:21:58.8 &  15:30:20 &      1,2,3,4,5,6,7 \\
04361+2547 & TMR 1  & 4:39:13.9 &  25:53:21 &      8,9,10         \\
04365+2535 & TMC 1A & 4:39:35.0 &  25:41:45 &      9,10,11,12,13  \\
04381+2540 &        & 4:41:12.5 &  25:46:37 &      9,10,11,12     \\
TMC 2A     &        & 4:31:55.9 &  24:32:49 &      1,14            \\
L1529      & TMC 2  & 4:32:44.7 &  24:25:13 &      1,15,16         \\
L1642      &        & 4:35:02.8 &  14:13:57 &      17              \\
\noalign{\smallskip}
\tableline
\noalign{\smallskip}
\multicolumn{5}{l}{\parbox{0.8\textwidth}{\footnotesize
1- \cite{lari99}; 2- \cite{fukui89}; 3- \cite{andre99};
4- \cite{lee02}; 5- \cite{bello02}; 6- \cite{taka03};
7- \cite{lee05}; 8- \cite{terebey90};
9- \cite{moriarty92}; 10- \cite{bontemps96};
11- \cite{terebey89}; 12- \cite{chan96};
13- \cite{tamura96}; 14- \cite{myers88};
15- \cite{lichten82}; 16- \cite{lada85}
17- \cite{lil89}
 }}\\

\end{tabular}
}
\end{center}
\end{table}

\section{Individual Objects of Interest}

\subsection{T Tauri}

\begin{figure}[p]
\plotone{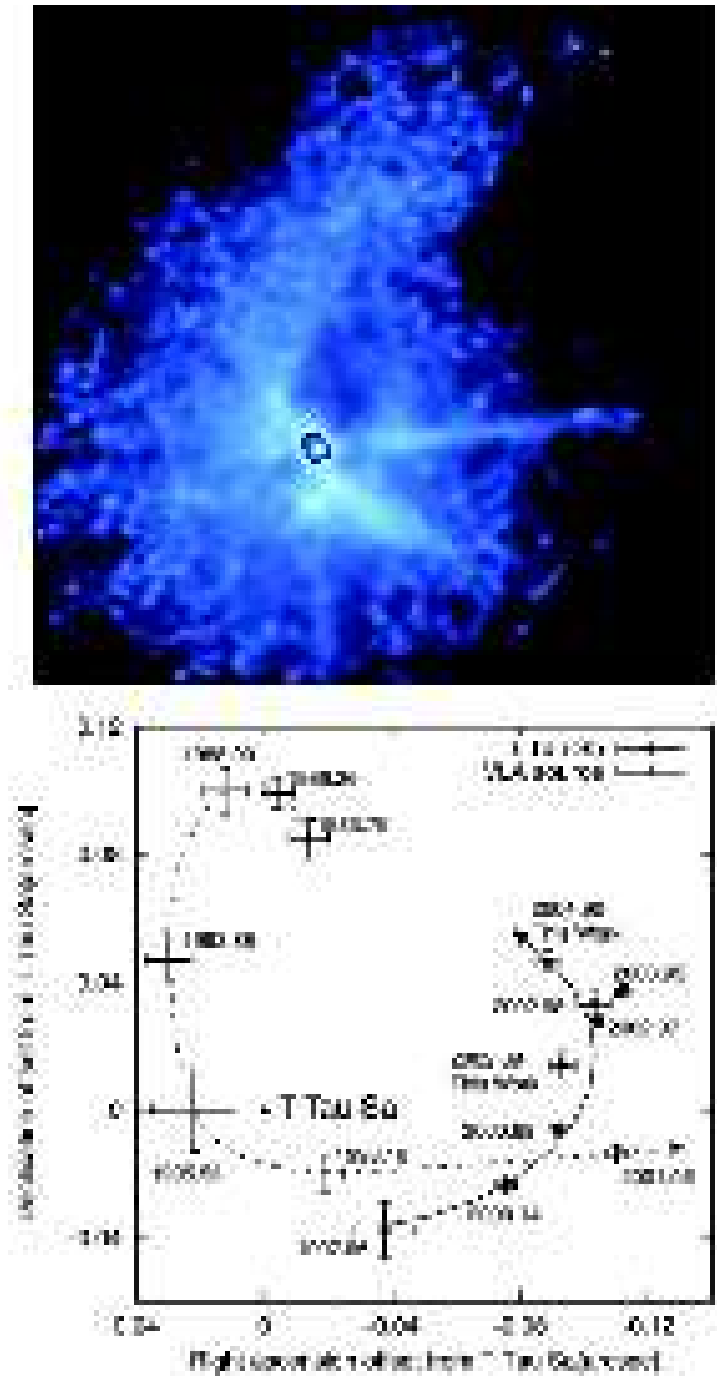}
\caption{The T Tauri binary. {\it Upper panel:}
10\arcsec x 10\arcsec~optical image (C. \& F. Roddier).
T Tau N is the saturated point source surrounded by a dark
halo. T Tau S is the fuzzy source due south. At PA = 315\deg~from
T Tau, the slightly elongated blob is a bright radio source
\citep{ray97}. Two collimated jets appear to emanate from T Tau N,
one with PA $\approx$ 45\deg~and 225\deg~and another with PA $\approx$
270\deg. The bright spike at PA $\approx$ 270\deg~points to Hind's
nebula, $\sim$ 45\arcsec~to the west.
{\it Lower panel:} Motion of the T Tau S binary from Mayama et al. (2006).
T Tau Sa and T Tau Sb appear to form a bound pair; the VLA source
is probably not bound to this pair \citep[see also][]{koh08}.
}
\label{fig:ttau}
\end{figure}

T Tauri is synonymous with young stars.  In addition to serving
as the prototype of low mass, young variable stars, T Tau has
inspired a festival for young filmmakers and countless images of
space art. Lying in rich nebulosity in the most southern of the
Taurus-Auriga dark clouds \citep{sta98}, T Tauri has an embedded
companion \citep[T Tau S;][]{dyc82,sch84} and a bright optical
jet \citep[Figure \ref{fig:ttau};][]{schwa75,buhr86,gor92,eislmun98}.

T Tauri has an amazing, 150 yr history of brightness variations
\citep[e.g.,][]{bax16,ryd84a,her94}.  The bright optical star,
T Tau N, has a K-type spectrum and fluctuates between V = 9
and V = 14 on timescales of months to years \citep{bec01a}.
There is also a small (0.01--0.03 mag) periodic variation in the
light curve due to stellar rotation \citep{her86}. Roughly 100--200 AU
away, T Tau S is a close binary composed of an intermediate mass star,
$\sim$ 2.1 $M_{\odot}$, and a lower mass pre-main sequence star with
a mass of $\sim$ 0.8 $M_{\odot}$ \citep{kor00,duc02,duc05,furl03,bec04,
john04,tam04,duc06,may06,koh08,ske08}.
During 1985--1993, the more massive component of T Tau S brightened
by more than two magnitudes in the IR and then faded
\citep{ghe91,kob94,sim96,bec01a,bec04}.
This behavior is reminiscent of FU Ori eruptions \citep{har96}.

Multiwavelength observations of the nebulosity reveal a complicated
structure \citep{sau03,may06}. The elliptical Burnham's nebula surrounds
T Tau and has an extent of a few arcsec.  Hind's nebula (NGC 1555) is
an arc-shaped reflection nebula 45\arcsec~to the west; Struve's small
nebula (NGC 1554) is 4\arcmin~west \citep{bur90,Bar95,Cur15,Her50}.
Recent IR, optical, and UV data show a patchy structure, with an
optical extinction of roughly 1-2 mag to T Tau N and $\sim$ 15--40 mag
to T Tau S \citep{kob97,vda99,bec01b,sau03}. Although early spectra
hinted at emission lines in Burnham's nebula, \citet{Her50} first showed
convincingly that the lines are intrinsic to the nebula. Deeper optical
spectra extended this study and demonstrated the lines are formed in
shocked gas \citep{sol88,boso94,vda99}.

Both T Tau N and T Tau S are bright radio sources
\citep{sch84,sch86,ray97} and drive powerful stellar winds
\citep{john03,john04,loi07a,bec08}. T Tau S also has a fairly
bright H$_2$O maser \citep{furu03}. A third bright radio source
roughly midway between the two stars is visible as a fuzzy blob
in the optical image (Figure \ref{fig:ttau}). A bright nonthermal
radio source associated with T Tau S provides an excellent distance
estimate of 147.6 $\pm$ 0.6 pc \citep{loi05,loi07b}. This emission
probes the structure of the magnetosphere and inner disk around the
brighter component of the T Tau S binary \citep{loi07a}.

Both T Tau N and T Tau S eject jets.  The jet in T Tau N was first
detected on photographs from the 1940s and 1950s \citep{Her50}.
Many spectroscopic and imaging studies identify shocked gas along
the well-collimated jet
\citep{schwa74,schwa75,boso94,boso99,bur94,eislmun98,wal03}. The brightest
part of this jet lies along PA $\approx$ 270\deg~and extends for at
least 30\arcsec, which is $\sim$ 5000 AU at 140 pc. A much fainter
counter-jet lies
along PA $\approx$ 90\deg. The kinematics of this weak flow are
consistent with kinematics of the brighter western jet~\citep{boso99}.

T Tau S drives a remarkable jet \citep{rei97,wal03}. Close to the star,
a broad, multicomponent bipolar jet (HH 255) lies along an axis with
PA $\approx$ 0\deg$\pm$30\deg \citep{sol88}. The jet has a length of
several arcsec, with a blueshifted southern lobe and a redshifted
northern lobe \citep{boso94,boso97,boso99,qui97}. At much larger
distances, \citet{rei97} identified a `giant' HH flow (HH 355) with
similar kinematics. For a distance of 140 pc, the HH 355 bipolar flow
has a projected length of 1.6 pc.

Circumstellar disks play an important role in the geometry of the
jets.  T Tau S has an extensive dusty disk with a mass of $\sim$
0.04 $M_{\odot}$ \citep{hog97b,ake98,wal03}. Material from this disk
probably drives the T Tau S jet and the large molecular outflow
\citep{edsne82,levreault88,moriarty92}. In T Tau N, 3D radiative
transfer calculations suggest that the rotational axis of the dusty
disk is tilted by $\sim$ 20\deg~relative to the rotational axis
in T Tau S \citep{wood01,ake02,ake05}.

\subsection{L1551 IRS 5}

L1551 IRS5 is the iconic bipolar outflow source. Although \citet{sharp59}
originally identified it as an H~II region, \citet{stro74, stro76} showed
that the object is a very red young star illuminating a reflection nebula.
More sensitive, multi-wavelength data demonstrate that the 0.5--1 pc long
outflow has several components. Bright optical and near-IR jets are
surrounded by a well-collimated molecular outflow embedded in a magnificent
reflection nebula \citep{snell80,snell85,snel85,dra85,sca88}.
The flow contains many HH objects and is a primary testing ground for
theories of jet formation and evolution.

\begin{figure}[htb]
\epsscale{1.0}
\plotone{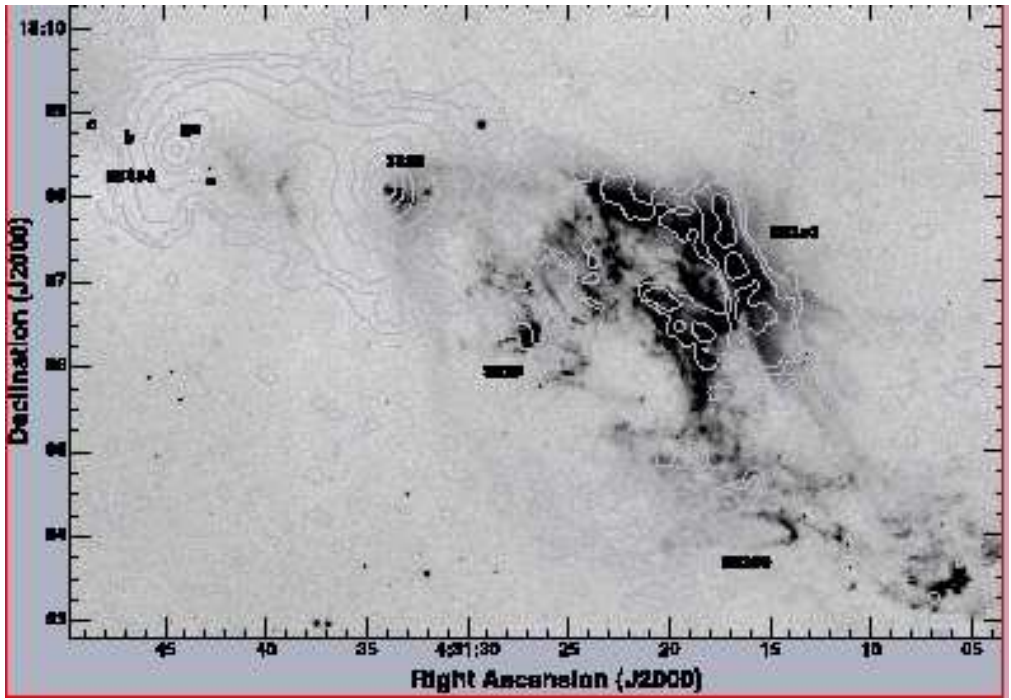}
\caption{
[S II] image of L1551 IRS5 with contours of 850 $\mu$m emission
overlaid (Moriarty-Schieven et al. 2006). The contour levels are
0.02, 0.04, 0.08, 0.16, 0.32, 0.64, 1.28, and 2.56 Jy beam$^{-1}$.
\label{fig:l1551}
}
\end{figure}

A relatively inconspicuous close binary drives the outflows in
L1551 IRS5 \citep{bei85,loon97,rod98,rod03}. The binary has a
projected separation of 45 AU, yielding a rough orbital period of 300 yr.
Both stars are bright radio sources at mm and cm wavelengths, with
thermal emission from circumstellar disks and free-free emission
from the jets and stellar winds. The disks lie in an extended envelope
that dominates the far-IR and submm flux \citep{bei81,dav84,eme84,edw86}.
The spectral energy distribution of scattered optical and near-IR
light and thermal far-IR and submm flux indicate that the extended
envelope is falling into the central object at a rate $\sim$ 3--10
$\times 10^{-6} M_{\odot} {\rm yr}^{-1}$ \citep[e.g.][]{ken93a,but94,whi97,
oso03,gram07}.

The structure of the IRS5 jet is remarkable (Figure \ref{fig:l1551}).
The flow consists of a short jet (HH 154) and numerous faint knots and
bow shocks
\citep[HH 28, HH 29, HH 256, HH 257, HH 259, HH 260, HH 261, HH 262,
HH 264, HH 265, HH 286, and HH 454; ][]{mufr83,stro86,sto88,grahe90,garnav92,dav95,hod95,krist98,lopez98}.  Both binary components appear
to drive outflows, which interact and merge to form the majestic
large-scale outflow structure \citep{fridli98,dev99b,har00,rod03}.
Some knots and bow shocks are associated with the nearby young stars
L1551NE and HH30 IRS \citep{herbig74,eme84,stro86,grahe90,garnav92,dev99b}.
Because there are at least three interacting HH flows, the kinematics
of the region is complicated, with some indication for twisted jets
and other distortions \citep{necsta87,camp88,sto88,yata92,frilis94,
fridli98,itoh00,fridl05,liseau05}. The HH objects require several
velocity components ranging from slow speeds of tens of km s$^{-1}$
up to velocities of 300 or more km s$^{-1}$ \citep{sto88,har00,rei00,fav02,pyo02,pyo05,bally03,dav03}.

In addition to the jet, L1551 IRS5 has a large-scale molecular
outflow \citep{snell80,bala83,moriarty87,sar88,pou91,moriarty92,
frikne93,bac94,ful95,lad95,pla95,yok03}.  The outflow is clumpy, with
significant small-scale structure, and a fast component in H~I
\citep{gio00}. The outflow rate of $\sim 10^{-5}$ $M_{\odot}$
yr$^{-1}$ is comparable to the infall rate \citep{oha96,sai96}.
Many of the small-scale features are associated with similar
structures in the jet and the HH objects. As in the optical
HH objects, the molecular gas shows evidence for shocks,
including X-ray emission \citep{rud92,bar93a,fav02,bally03}.

The jet is embedded in an impressive infalling cloud of molecular
gas \citep{lay94,oha96,sai96,momose98,tak04a}. The kinematics
of the gas indicates infalling material with a central mass of
$\sim$ 0.1--0.15 $M_{\odot}$, consistent with the mass estimate
derived from the optical luminosity \citep{ken99} but much smaller
than estimated from the binary orbit, $\sim$ 1.2 $M_{\odot}$
\citep{rod03}. The infall rate is $\sim 5 \times 10^{-6}$
$M_{\odot}$ yr$^{-1}$, consistent with rates derived from the
spectral energy distribution.

A striking reflection nebula yields additional information about
the central object, the outflowing jet, and the infalling gas.
Optical spectra indicate that at least one component of the central
binary is an FU Ori object \citep{mun85,car87,sto88,sand01}. Near-IR
data show that this component is variable on short timescales \citep{liu96}.
The nebulae is highly polarized, $p \sim$ 15\%--20\% at JHK, with
vectors perpendicular to the outflow axis \citep{luc96,whi97}. These
data demonstrate that the nebula marks the boundary of a cavity produced
where the jet flows out through the collapsing cloud.

L1551 IRS5 provides excellent tests for numerical models of star
formation.  In the current picture, gas in a molecular cloud core
collapses into a star + disk system, where the central star slowly
grows from material transported inward by the disk. Calculations of
the collapse yield predicted images and spectral energy distributions
for comparison with observations \citep[e.g.,][]{ada86,ada87,ada88}.
In L1551 IRS5, models with a flattened infalling envelope surrounding
a pair of circumstellar disks account for the images, polarization
maps, and spectral energy distribution \citep{str85,but91,keene90,ken93a,
ken93b,whi00,oso03}.

\subsection{RW Aurigae}

Discovered as an irregular variable \citep{cer06} with an optical
companion at a separation of 1.5 arcsec \citep{joy44,ghe93,ghe97,whi01},
RW Aur may be a spectroscopic binary \citep{Gahm99,pet01}.  In
addition to the strong red continuum of a typical T Tauri star,
the system has a strong UV continuum, hundreds of strong emission
lines, and a prominent near-IR excess \citep{joy45,Herb45,
men66,gla74,sha79}. Because absorption lines from the underlying
star are difficult to detect on most spectra, RW Aur and other
T Tauri stars with similar spectra are often called `continuum
+ emission' stars.

\begin{figure}[tb]
\centering
\plotone{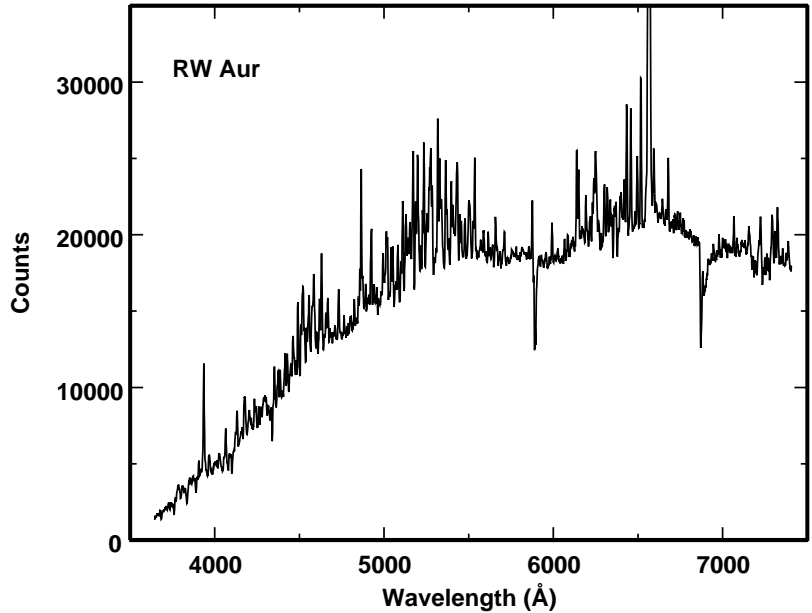}
\vskip -10ex
\caption{Optical spectrum of RW Aurigae from \citet{ken98}.
The spectrum shows a strong red continuum from a late-type
star along with a weak blue continuum and strong emission
lines from a hotter, optically thick source. Aside from
the strong H~I Balmer lines, the spectrum contains emission
from [O~I], [S~II], Fe~II, and [Fe~II].
}
\label{fig:rwaur}
\end{figure}

RW Aur is among the most active T Tauri stars. The optical continuum
and strong line spectrum (Figure \ref{fig:rwaur}) vary irregularly on
timescales of hours to months; there is a small amplitude variation
with an underlying periodicity of 2.6--2.8 days
\citep{Herb45,Gahm70,app83,Ivan93,Gahm99,pet01}. As the system brightens,
the optical and near-IR colors become bluer; as the system fades,
the colors become redder. Together with analyses of the optical
spectrum, this behavior indicates a high accretion rate from the
disk onto the central star.

The spectrum of RW Aurigae is distinct from most T Tauri stars.
At low resolution, the absorption lines are almost completely
veiled by a strong, blue continuum and many emission lines.
At higher spectral resolution, the line ratios of temperature
sensitive absorption lines imply a K7 or M0 spectral type,
much cooler than the middle G spectral type estimated from
early spectra \citep{hart89,hart91}.
In the UV, a strong Balmer continuum demonstrates that dense gas
with a temperature of roughly 10,000 K produces the blue excess
\citep{imh80,err00}.
Strong line emission from [C IV] and H$_2$ indicates a wide range
of temperatures.

RW Aur also has a remarkable bipolar jet
\citep[HH~229; ][]{baccio96,woi02b,woi05,lopez03,gomezdec03,ale05,pyo06,bec08}.
The data indicate a well-collimated outflow
from a small inner region of the disk, with variations on timescales
of months to years. The jet may rotate close to the star, with high
velocities in the C~III] and Si~III] lines suggestive of a rotating
belt or ring close to the star\footnote{Coffey et al. (2007) describe
evidence for rotation in the micro-jet of DG Tau.}. At larger distances,
the blue and red lobes of the outflow appear to have different helicities,
as predicted by MHD theories. It is not clear whether changes in the jet
are associated with fluctuations in the brightness of the underlying star
and disk. With some evidence for a 20 yr period in the outflow rate,
searches for similar timescales in the optical source might provide
additional tests of jet models.

The structure of molecular gas around RW Aur is also interesting.
RW Aur A lies within a small molecular disk (40--60 AU radius;
disk mass $\sim 3 \times 10^{-4}$, Andrews \& Williams 2005) in
Keplerian rotation about the central star and oriented perpendicular
to the bipolar jet \citep{cab06}. RW Aur B is within an asymmetric
clump of gas connected to the RW Aur A disk by a 600 AU arm of
material between the two stars. \citet{cab06} interpret this arm
of gas as a tidal tail resulting from the orbital interaction of
RW Aur A and B.

\subsection{RY Tauri}

RY Tau is embedded in an impressive reflection nebula
\citep{herbig61,petrov99}. The central star and the nebula are
variable on hour to year timescales. Although emission lines
from the nebula have been observed on several occasions, recent
observations demonstrate a 31\arcsec ($\sim$ 4500 AU) jet in
H$\alpha$ \citep[][see also G\'omez de Castro \& Verdugo 2007]{sto08}.
The orientation of the jet on the sky is roughly perpendicular
to the plane of the disk derived from polarization measurements.
The young dynamical age of the jet system, short timescale optical
variability, and strong intercombination emission lines suggest that
this system might yield interesting clues concerning the structure
of jets and the inner accretion disk \citep[see also][]{sche08}.

\begin{figure}[htb]
\centering
\epsscale{0.85}
\plotone{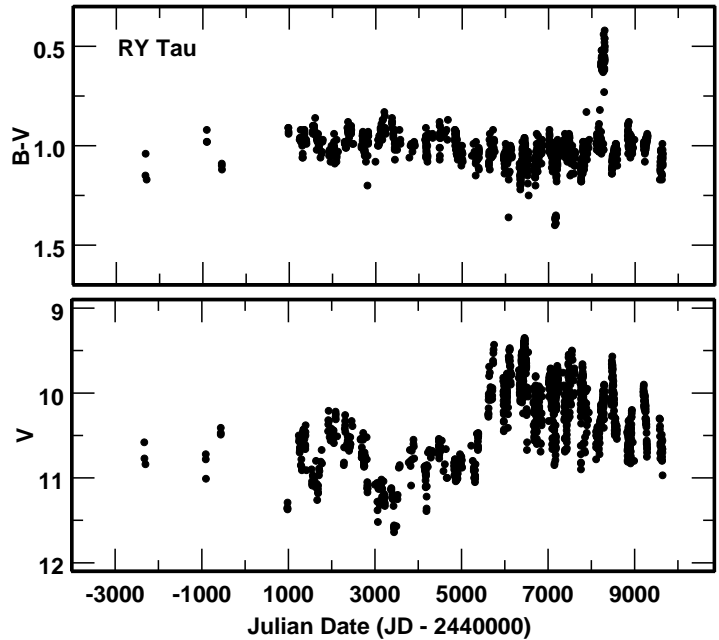}
\vskip -1ex
\caption{Optical light curve of RY Tauri using data from
the literature \citep{her94}. The brightness and color
vary irregularly on various timescales.}
\label{fig:rytau}
\end{figure}

The continuum and emission lines in RY Tau vary on timescales
of hours to months with no obvious periodicity
\citep[Figure \ref{fig:rytau};][]{drago71,zajt74,zajt80,zajt82,
zajt85,herbst84,holtz86,ismail87,chu91,vrba93,bec01a}. On long
timescales, the star usually becomes bluer when brighter and
redder when fainter. Although the equivalent width of the H$\alpha$
emission line
often increases as the star fades, the absolute H$\alpha$ flux declines.
When the star is faint, large flares on hour timescales are common.
The flares have blue colors, which requires a hot source.  Occasionally,
the star becomes bluer when it brightens and redder when it fades,
indicating that variable extinction along the line of sight causes
some fluctuations.

RY Tau is polarized, $p \sim$ 2\%--3\%, with variations on timescales
similar to the fluctuations in the optical continuum level
\citep{vard64,efi80,bast82}. The magnitude of the polarization shows
that the nebula partially obscures the central star. From high quality
spectropolarimetry, \citet{vin03} suggest that much of the H$\alpha$
emission is scattered off a rotating circumstellar disk
\citep[see also][]{koe95}.

\subsection{DR Tauri}

DR Tauri is another `continuum + emission' T Tauri star with
a rich emission line spectrum. It is a member of the small
group of `EXors,' T Tauri stars that have brightened by 2--4
mag and remained bright for months to decades \citep{herbig89}.
As in most T Tauri stars, there is a strong UV excess from
Balmer continuum emission and a large IR excess from material
in the disk \citep{gue93,ken94,skr96,hes97}.

The recent outburst of DR Tau is the slowest-developing of the
EXor class. Prior to 1960, the star maintained B $\approx$
14 for $\sim$ 30 yr \citep{cha79,gotz80a,gotz80b,kur80}.
From 1960--80, the star brightened by $\sim$ 3.5 mag. As the
system brightened, the colors became bluer.
Since 1980, the star has remained bright with no evidence
for fading as in some FUors.

The spectrum of DR Tau is highly variable. The emission lines
and blue continuum have night to night variations of 0.5 mag
and smaller variations on hour timescales \citep{ber77,kra80,
kol87,app88,gue93,ken94,hes97}.  There is some evidence for 5 day
and 10 day periodicities in the continuum and the lines. These
variations may reflect the underlying stellar rotation period.
The IR continuum also varies \citep{ken94,skr96}, suggesting
strong correlations in activity between the hot source and the
cool source.

Because the variations in the lines and continuum are so well-correlated,
DR Tau is one of the best examples of magnetospheric accretion among
T Tauri stars \citep{gue93,ken94,hes97,smi97,smi99,berist98,mah98}. In
this picture, a circumstellar disk produces the strong IR excess;
changes in disk luminosity produce variations in the magnitude of
the excess. A strong magnetic field truncates the disk several
stellar radii above the stellar photosphere and channels material
onto the star. Material in the stream is heated from roughly 1000 K
(the temperature of the inner disk) to roughly 10,000 K. Shocked gas
where the stream hits the star produces a wide range of emission lines.
The large tilt of the magnetic axis with respect to the rotation axis
produces
variations in the UV continuum and line emission, including the apparent
occultation of material in the stream or in the shock as these structures
rotate behind the stellar photosphere.

Although a jet has not been discovered in DR Tau, the line profiles
show evidence for accretion and outflow \citep{app80,ale01,ard02}.
Deep narrow-band images might reveal whether the system has a jet.

\subsection{HL Tauri}

HL Tauri is a remarkable T Tauri star in a complex region
of star formation. In the early 1980's, near-IR and mid-IR
observations suggested an edge-on circumstellar disk surrounding
a highly variable T Tauri star \citep{cohen83,gras84,bec86}.
High resolution radio observations supported this interpretation;
later observations indicated the disk lies at the center of a
large-scale molecular inflow and outflow \citep{sarbec87,beck89,
gras89,monin89,sar91,rod92,hay93,lay94,lay97,cabrit96,wil96,brit05}.
The disk is geometrically flared and massive, with a radial
temperature gradient close to that predicted by theory.

\begin{figure}[!ht]
\centering
\epsscale{0.80}
\plotone{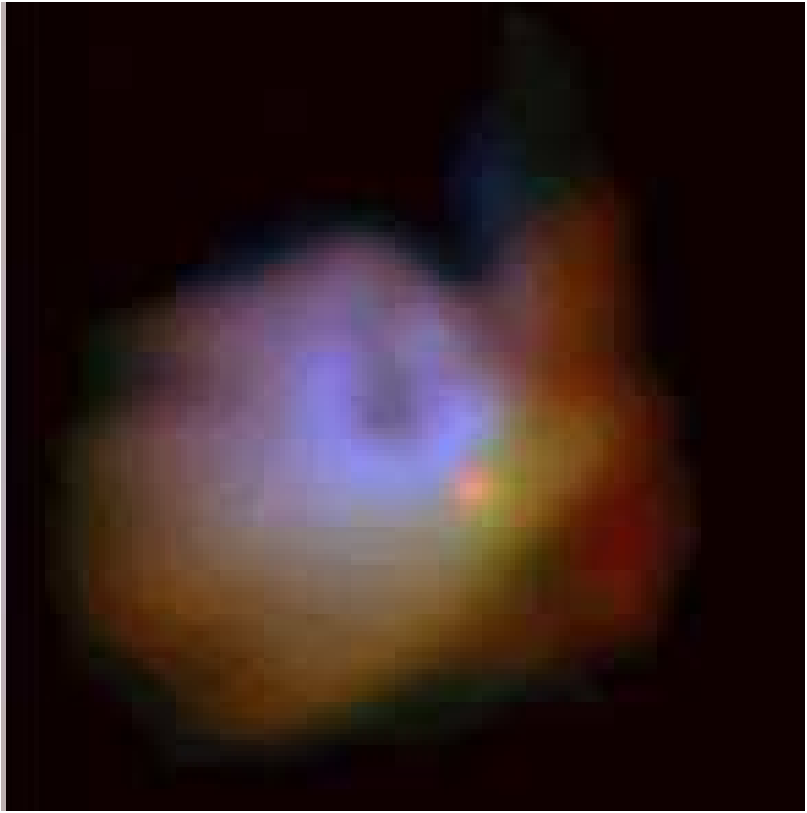}
\caption{Image of HL Tauri in the I, J, and H filters from
\citet{close97}. The image is roughly 6$^{\prime \prime}$
on a side. Blue colors (I) trace material close to the
outflow cavity; green colors (J) trace material in the disk
and infalling envelope; red colors (H) trace scattered emission
from the star and highly reddened light in the disk.
}
\label{fig:hltau}
\end{figure}

HL Tau is heavily embedded within an impressive C-shaped reflection
nebula \citep[Figure \ref{fig:hltau};][]{sta95,wein95,close97,welch00}.
On large scales, this nebula merges with nebulosity surrounding XZ Tau.
In HL Tau,
the star and the nebula are variable and highly polarized, indicating that
the star is observed only in reflected light \citep{wein95,liu96,whi97,lucas04}.
Optical and near-IR spectra show a heavily veiled, K7-M0 spectral type,
with strong emission from H~I and various metallic lines
\citep[e.g.,][]{coh79,carr89,carr90,ken95,ken98}.

In addition to the nebula, HL Tau drives a bright optical jet (HH 150) and
an energetic molecular outflow \citep{mufr83,cohejo87,torre87,mundt87,
mundt88, mundt90,magak89,rod94,monin96,mun96,lopez95,lopez96}. This
region has additional outflows from HH30 IRS, XZ Tau, and the optically
invisible object HL Tau VLA-1 northeast of HL Tau
\citep{mundt87,mundt88,mundt91,mov07}, which has complicated interpretations
of the jet structure and driving mechanism. The nearby late-type T Tauri
star LkH$\alpha$ 358 does not appear to drive an outflow, but deeper
images might reveal more structure in this fascinating region.

Close to the star, there is a well-collimated micro-jet in [Fe~II] with
clear bipolar morphology \citep{pyo06,tak07}. Both components of the
micro-jet lie within bubbles of H$_2$ emission \citep[see also][]{bec08}.
The length of the micro-jet,
$\sim$ 150 AU, is comparable to the diameter of the bubble.  The NE H$_2$
bubble lies within the scattered light emission in Figure \ref{fig:hltau}).
The much fainter SW component may lie within a much fainter bubble with
similar structure.

\subsection{Haro 6-10}

Haro 6-10 is a relatively nondescript young binary star driving a remarkable
giant HH flow. After \citet{haro53} discovered the Haro 6-10 emission knot,
\citet{elias78} detected a bright source with an IR excess at the base of
a small fan-shaped nebula.  The central binary has a separation of $\sim$
1.2 arcsec \citep{lei89,men93,kor99,whi01,duc04}. The primary (Haro 6-10 S)
has a late spectral type \citep{good86,carr90,her95} and may be a binary
\citep{rei04}; the secondary (Haro 6-10 N) may be a continuum + emission
star.  The system also has a large far-IR excess and lies embedded in a
dense envelope in a small ammonia core \citep{ang89,sat90}. Haro 6-10 N
appears to produce most of the far-IR excess.

The Haro 6-10 outflow consists of several small jets, a few HH knots, and
a giant outflow with a length of 1.6 pc \citep{stro86,dev99a,mov99,rei04}.
The original HH 184 is composed of a small set of knots subtending a wide
opening angle.  These knots are probably part of two independent flows
ejected by each component of the binary. On small scales, Haro 6-10 S drives
an optical jet \citep{mov99} and a compact radio jet \citep{rei04}.
On large scales, HH 410 and HH 411 mark the ends of the outflow. A few
other knots, including HH 412, lie between the binary and the end of the
outflow \citep{dev99a}.

Haro 6-10 is also an active mm and submm radio source.  The system has a
large-scale molecular outflow that is roughly aligned along the direction
of the optical outflow \citep{terebey89,dev99a,sto07}.  Strong extended
submm continuum emission from Haro 6-10 lies perpendicular to the outflow
direction \citep{cha98,mott01}. There is some indication that the extended
envelope is still infalling into the disk of Haro 6-10 \citep{cha98}.

Like T Tau and L1551 IRS5, the Haro 6-10 binary varies significantly at near-IR
wavelengths \citep{lei01}. Some of the variations -- including changes in the
equivalent width of the 3.1 $\mu$m ice absorption band -- are consistent
with fluctuations in the line-of-sight extinction. However, some variations
are probably associated with the binary components. Long-term monitoring of
near-IR variability in the brightest Taurus-Auriga sources would provide a
context for interpreting the variations in the more famous sources such as
T Tau, L1551 IRS5, and Haro 6-10.

\subsection{IC 2087}

\begin{figure}[!ht]
\plotone{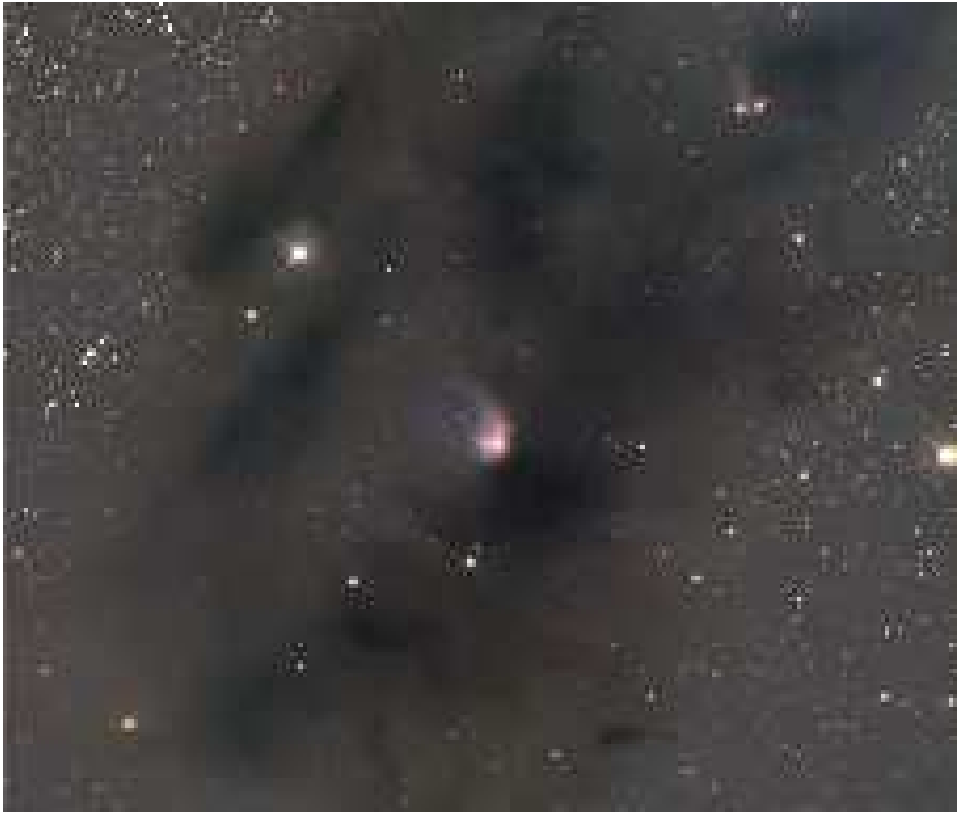}
\caption{Optical image of IC 2087 and surroundings (courtesy Thomas V.
Davis).  The image is roughly 1 degree on a side; north is up and east
is to the left. A heavily reddened T Tauri star (IC 2087 IRS) and HH 395
lie within the bright reflection nebula at the center of this image.
The dark areas are regions of high extinction within the B22 and L1527
dark clouds.  Several protostars -- IRAS04361+2547, IRAS04365+2535, and
L1527 IRS -- and at least one HH object -- HH 192 -- lie within the small
dark clouds north and west of IC 2087. The bright pair of nebulous stars
in the SE corner is a small group of binary T Tauri stars (V955 Tau,
LkH$\alpha$ 332/G1, and LkH$\alpha$ 332/G2). The bright set of nebulous
stars in the NW corner is another group of T Tauri stars (DO Tau, GM Tau,
and the HV Tau triple system). These stars power several HH objects,
including HH 230 and HH 831--834.  Other fainter T Tauri stars and HH
objects are scattered across the field.
}
\label{fig:ic2087}
\end{figure}

Located in the so-called Taurus Molecular Ring
\citep[TMR; e.g.,][]{schl84,cer87,terebey90,toth04},
IC 2087 is a small, bright reflection nebula
\citep{hod94} embedded in a dense ammonia core associated with L1534
\citep{ben89}, see Fig.~\ref{fig:ic2087}. The bow-shaped nebula surrounds a very red T Tauri star
\citep[IC 2087 IRS; ][]{allen72,elias78}, associated with a molecular outflow
\citep{heyer87}. The source is a bright IRAS source, with a Class II mid-IR
to far-IR spectral energy distribution
\citep{bei86,harr88,berr89}. Although initial work assigned an early B
spectral type to the star, more recent spectra and the integrated
luminosity suggest a highly reddened K4 PMS star
\citep{ken95,whi04}. The K4 spectrum is heavily veiled, with strong
H$\alpha$ and Ca~II emission but no obvious forbidden lines.

In addition to driving the molecular outflow, the central TTS is linked
to two HH knots well outside the main body of the reflection nebula
\citep[HH 395A and HH 395B; ][]{gomez97}. The knots are bright in [S~II]
but are not visible in H$_2$. A line connecting the two knots points
back to a small indentation in the highly polarized reflection nebula
\citep[$\sim$ 2\%--4\% at JHK;][]{tam89,whi97} and the bright IR source.

The line-of-sight towards IC 2087 has been a popular laboratory for
studying ice formation in molecular clouds \citep[e.g.][]{sat90,tie91,
chiar95,bro96,bow98,tei99,shup01}. As in other regions of Taurus-Auriga,
H$_2$O ices and CO/CO$_2$ ices are detected along various sightlines through
the cloud. The data indicate that the optical depth towards the TTS is
larger than the optical depth through the cloud, which requires additional
absorbing material along the sightline to the IR source.

\subsection{VY Tau}

VY Tau is an exceptional PMS star in the Taurus-Auriga dark clouds.
The star is a close binary with a projected angular separation of
0.66" and approximate component masses of 0.6 $M_{\odot}$ and 0.25
$M_{\odot}$ \citep[e.g.,][]{whi01,woi01}. The primary star has an
M0 spectral type and is not veiled \citep{har90,shib93,val93}. The
system has no IR excess and is not a mm source \citep{ken95,oster95}.
However, the system is a bright X-ray source \citep{neu95}, with a
modest rotational velocity of $\sim$ 10 km s$^{-1}$ \citep{har89}.

Unusual optical variability and spectroscopic activity distinguish
VY Tau from other classical and weak-emission T Tauri stars
\citep{mein69,mein71,mein80,her77,herbig90,herb83,stone83,ryd84a}.
During decade-long periods of inactivity, the star is relatively faint,
B $\approx$ 15, and the M0 absorption spectrum is conspicuous.
The star then displays a fairly prominent 5.37 day period in its optical
light curve, which is consistent with its expected radius and rotational
velocity \citep{bou95}.  During occasional active periods of 1--2 yr,
the star brightens to B = 11-12 and develops an impressive emission
spectrum with bright, low excitation lines from Fe~I, Si~I, Mg~I. Although
higher excitation lines from Fe~II and the H~I Balmer series are often
present, they are weak. Both sets of features seem to vary at roughly
constant equivalent width as the star varies between B = 11--14.

\citet{herbig90} discussed several possible interpretations for the
origin of the activity in VY Tau. With the discovery of a close
companion, the two binary hypotheses currently seem most worthy of
further study. If the fainter component of the system is surrounded
by an optically thick ring or disk of dust, active periods might occur
when the optical depth in the dust declined. Although close approaches
of the two stars might induce active periods, the wide separation --
$\sim$ 90 AU -- probably precludes activity cycles with recurrence
times of 1--2 decades.  If deep mid-IR and submm studies of this system
detect evidence for emission from a dusty ring or disk, detailed
study of this emission might reveal clues to the origin of the amazing
line spectrum.

\subsection{L1527 IRS}

Lying within a dense ammonia core \citep{ben89}, L1527 IRS is the most
deeply embedded protostellar source in the Taurus-Auriga dark clouds
\citep[e.g.][]{ken90b,ken93a,ken93b,and93,ken95,chin01,whi97,har05,furl08}.
Several groups have classified this system as a Class 0 source, implying
that it lies at an earlier evolutionary stage than typical Class I sources
\citep[e.g.,][]{and93,chin01}. Detection of C$_4$H$^-$ and other
carbon-chain molecules suggests the nebula has interesting chemistry
\citep{sak07,agu08,sak08a,sak08b,sak08c,has08}.  Despite its relatively
low luminosity of $\sim$ 1 $L_{\odot}$, the central young binary star
drives a wide-angle molecular outflow, illuminates an enormous variable,
bipolar reflection nebula, and powers a fascinating set of HH objects, HH 192.

\begin{figure}[!ht]
\plotone{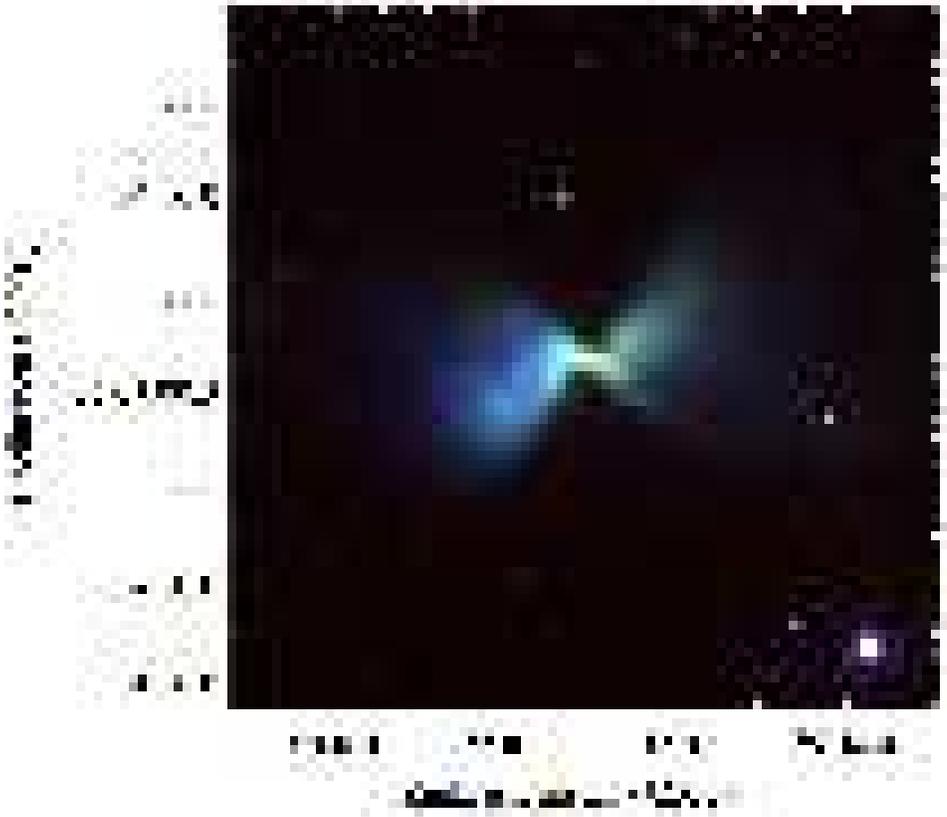}
\caption{Spitzer image of L1527 IRS in the [3.6], [4.5], and [5.8] filters
from the Taurus Legacy program \citep[see][]{gud07a}.
Blue light ([3.6]) traces scattered light far away from the central
protostar; green light ([4.5]) and red light ([5.8]) traces scattered
light from material closer and closer to the protostar.
}
\label{fig:l1527}
\end{figure}

At optical and near-IR wavelengths, L1527 IRS displays an almost
perfectly symmetric bipolar reflection nebula with an angular size
of more than 1 arcmin
\citep[Figure \ref{fig:l1527};][]{ken93b,eiroa94,tamura96,whi97}.
The eastern lobe of the nebula is much brighter than the western
lobe and is highly polarized, with $p_K \sim$ 40\% to 60\%. This
lobe shows strong evidence for a large-scale outflow, with strong
H$\alpha$, [O~I], and [S~II] emission lines on optical spectra
\citep{eiroa94,ken98} and two amorphous [S~II] knots (HH 192A and
HH 192B) on deep images \citep{gomez97}.  A single small HH object (HH 192 C)
lies several arcmin west of the reflection nebula. Deeper images on larger
telescopes might detect additional optical jet emission from the western
lobe of the nebula.

In addition to the HH objects, the bipolar nebula marks an impressive
molecular outflow which lies perpendicular to an extended envelope of
infalling gas \citep{mye95,ful96,zho96,bontemps96,tamura96,hoger97a,hog98,
chin01,loi02,rei04}. The outflow lies in the plane of the sky and has a
large opening angle of $\sim$ 50\deg. A binary cm radio source coincides
with the position of the central, bright IRAS source (04368+2557); this
region powers one or two outflows observed at cm and mm wavelengths as
well as the optical HH objects.

As with many of the Class I sources in Taurus-Auriga, the evolutionary
status of the L1527 IRS protostar in unclear. The massive infalling
envelope, $\sim$ 0.5--1 $M_{\odot}$, and relatively low luminosity
imply a very young central star with only a small fraction of its
`final' stellar mass (Class 0 source). The geometry of the molecular
outflow and the bipolar reflection nebula suggests that the central
source lies embedded in an extended edge-on disk. Extinction from the
disk and the infalling envelope produce the very red SED of the IRAS
source \citep{ken93a,whi97,and05}; scattering from the outflow
cavities produces the blue images and SED at IRAC wavelengths
(Whitney et al. 2003, Robitaille et al. 2007; Tobin et al. 2008).
This geometry complicates direct measurements of the luminosity
and the evolutionary status of the central young star.


\vspace{0.5cm}

\acknowledgements
B. Reipurth deserves special thanks for identifying the
beautiful images of T Tau (Figure 1) and IC 2087 (Figure 14),
his many incisive comments on the manuscript, and his unflagging
dedication to and support for star formation research.  We also
thank M. Geller, P. Goldsmith, and K. Luhman for detailed
and helpful comments on the manuscript, D. Goldman for his amazing image
of T~Tau in Figure 1,
K. Luhman for extraordinary help in verifying
the list of TTS in Table 3,
S. Mayama for the line drawing in Figure 9,
G. Moriarty-Schieven for Figure 10, Thomas V. Davis for his gorgeous image
of IC 2087 in Figure 14, and A. Accomazzi for help with technical issues.
We used the SIMBAD database, operated at CDS, Strasbourg, France,
and NASA's Astrophysics Data System Bibliographic Services.
This publication uses data products from the Spitzer Space Telescope
dowloaded from the Spitzer Science Center, the Two Micron All Sky Survey,
which is a joint project of the University of Massachusetts and the
Infrared Processing and Analysis Center/California Institute of
Technology, funded by the National Aeronautics and Space Administration
and the National Science Foundation, and the USNO-B catalog downloaded
from the United States Naval Observatory website.  We also used the
NASA/IPAC Infrared Science Archive, which is operated by the Jet
Propulsion Laboratory, California Institute of Technology, under
contract with the National Aeronautics and Space Administration.
The {\it NASA} {\it Astrophysics Theory Program} (grant NAG5-13278)
and the {\it Spitzer Guest Observer Program} (grant 20132) supported
portions of this project.


\end{document}